\documentclass{article}

\usepackage{PRIMEarxiv}

\usepackage{algorithm}
\usepackage{algpseudocode}
\usepackage{glossaries}
\usepackage[numbers]{natbib}
\usepackage{amsmath, amsfonts, amssymb}
\usepackage{mathtools}
\usepackage{booktabs}
\usepackage{threeparttable}
\usepackage{siunitx}
\usepackage{makecell}
\usepackage{array}
\usepackage{subcaption}
\usepackage{siunitx}
\usepackage{multirow, multicol}
\usepackage{soul}

\title{Neural Corrective Machine Unranking
\thanks{\textit{\underline{Corresponding author}}: 
\textbf{Georgina Cosma}} 
}

\author{
  Jingrui Hou \\
  School of Information Management,
  Wuhan University,
  Wuhan, Hubei, China \\
  Department of Computer Science,
  Loughborough University,
  Loughborough, Leicestershire, UK \\
  \texttt{houjingrui@whu.edu.cn} \\
  \AND
  Axel Finke \\
  Department of Mathematical Sciences \\
  Loughborough University \\
  Loughborough, Leicestershire, UK \\
  \texttt{a.finke@lboro.ac.uk} \\
  \And
  Georgina Cosma \\
  Department of Computer Science \\
  Loughborough University \\
  Loughborough, Leicestershire, UK \\
  \texttt{g.cosma@lboro.ac.uk} \\
}
\usepackage{xcolor}
\usepackage[normalem]{ulem}
\usepackage{cancel}

\newcommand{\jerryRevise}[1]{{\textcolor{black}{#1}}}

\newcommand{\querySet}{Q}
\newcommand{\documentSet}{D}

\newcommand{\documentSpace}{\mathcal{\documentSet}}
\newcommand{\querySpace}{\mathcal{\querySet}}
\newcommand{\datasetSpace}{\mathcal{S}}
\newcommand{\parameterSpace}{\mathcal{W}}

\newcommand{\trainedModel}{\mathcal{M}_{\mathrm{train}}}

\newcommand{\correctiveModel}{\mathcal{M}_{\mathrm{correct}}}

\newcommand{\originalSet}{\ensuremath{\text{\textsf{\upshape{S}}}}}

\newcommand{\forgetSet}{\ensuremath{\text{\textsf{\upshape{F}}}}}
\newcommand{\retainSet}{\ensuremath{\text{\upshape{\textsf{R}}}}}

\newcommand{\singleQuery}{q}
\newcommand{\singleDocument}{d}

\newcommand{\hingeLoss}{\mathrm{H}}
\newcommand{\quantile}{\mathrm{qtl}}
\newcommand{\rank}{\mathrm{rank}}

\newacronym{nlp}{NLP}{Natural Language Processing}
\newacronym{ir}{IR}{information retrieval}
\newacronym{nir}{NIR}{neural information retrieval}
\newacronym{dr}{DR}{dense retrieval}
\newacronym{nr}{NR}{neural ranking}
\newacronym{llm}{LLM}{large language model}
\newacronym{plm}{PLM}{pretrained language model}
\newacronym{cocol}{CoCoL}{Contrastive and Consistent Loss}
\newacronym{curd}{CuRD}{Corrective unRanking Distillation}
\newacronym{mrr}{MRR}{Mean Reciprocal Rank}

\newcommand{\loss}{\lossItem}
\newcommand{\lossItem}{\ell}

\newcommand{\reals}{\mathbb{R}}
\newglossaryentry{bertcat}
{
    name=BERTcat,
    description={BERT for concatenated query-document scoring}
}

\newglossaryentry{bertdot}
{
    name=BERTdot,
    description={BERT with Dot Production for separated query-document scoring}
}

\newglossaryentry{colbert}
{
    name=ColBERT,
    description={Contextualised Late Interaction over BERT}
}

\newglossaryentry{parade}
{
    name=PARADE,
    description={Passage representation aggregation for document reranking}
}

\usepackage{url}
\usepackage{hyperref}%
\hypersetup{%
  breaklinks=true,%
  linktocpage=true,% only link pagenumbers in toc
  colorlinks=true,%
  linkcolor=blue,%
  citecolor=blue,%
  filecolor=blue,%
  urlcolor=blue,%
  pdftitle={},%
  pdfsubject={},%
  pdfauthor={},%
  pdfkeywords={},%
  pdfproducer={}%
  }%

\begin{document}
\maketitle

\begin{abstract}
Machine unlearning in neural information retrieval (IR) systems requires removing specific data whilst maintaining model performance. Applying existing machine unlearning methods to IR may compromise retrieval effectiveness or inadvertently expose unlearning actions due to the removal of particular items from the retrieved results presented to users. 
We formalise \emph{corrective unranking}, which extends machine unlearning in (neural) IR context by integrating substitute documents to preserve ranking integrity, 
and propose a novel teacher--student framework, Corrective unRanking Distillation (CuRD), for this task.  
CuRD (1) facilitates forgetting by adjusting the (trained) neural IR model such that its output relevance scores of 
to-be-forgotten samples mimic those of low-ranking, non-retrievable samples; (2) enables correction by fine-tuning the relevance scores for the substitute samples to match those of corresponding to-be-forgotten samples closely; (3) seeks to preserve performance on samples that are not targeted for forgetting. We evaluate CuRD on four neural IR models (BERTcat, BERTdot, ColBERT, PARADE) using MS MARCO and TREC CAR datasets. Experiments with forget set sizes from 1\,\% and 20\,\% of the training dataset demonstrate that CuRD outperforms seven state-of-the-art baselines in terms of forgetting and correction while maintaining model retention and generalisation capabilities.
\end{abstract}

% keywords can be removed

\section{Introduction} \label{sec:intro}
\glsresetall

Machine unlearning refers to removing specific data from a trained machine-learning model while retaining its overall performance \citep{bourtoule2021machine,xu2023machine}. Machine unlearning presents unique challenges in the domain of \emph{\gls{ir}} \citep{Zhang2022}, and the application of unlearning to \gls{ir} is an area that is still largely under-explored.

\begin{figure*}[ht]
\centering
\includegraphics[width=\textwidth]{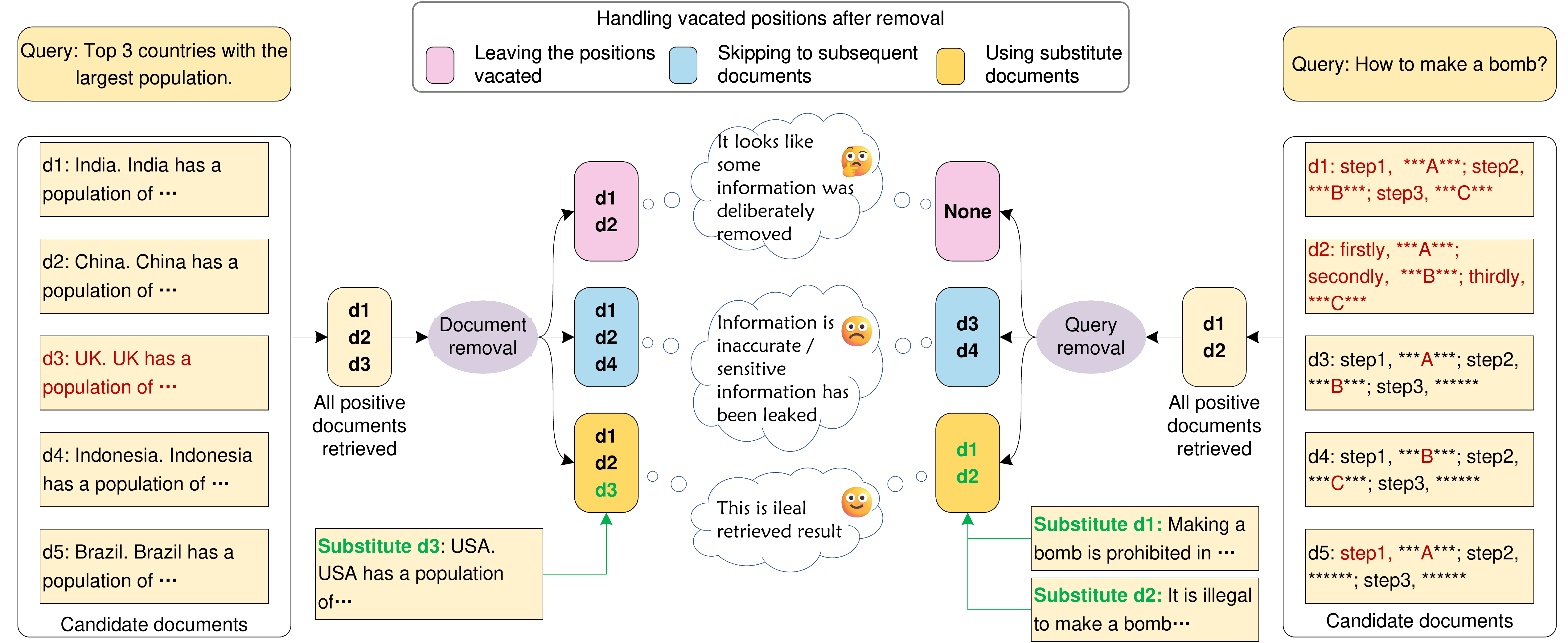} 
\caption{Illustration of the risks associated with data point removal in information retrieval and the mitigating effects of corrective unranking. Removing document `d3' results in problematic outcomes: if `d1' and `d2' are shown alone, users may perceive information deletion; if shown with `d4', it can lead to misinformation. Similarly, removing queries that retrieve correct answers `d1' and `d2' leads to either an empty result list, enhancing the perception of information deletion, or the presentation of `d3' and `d4', which may also inadvertently leak sensitive information. Employing substitute documents as a strategy in corrective unranking helps address these issues.} 
\label{fig:task.comparison}
\end{figure*}

In neural \gls{ir} models, unlearning may entail \textit{document removal,} i.e., making the model selectively forget certain retrieved documents; or \textit{query removal,} i.e., erasing specific query knowledge from the model.  \jerryRevise{It is important to emphasize that, in \gls{ir} context, unlearning entails reducing the learned association strength for selected query-document pairs rather than removing entire query or document entities. This typically involves modifying continuous relevance scores or adjusting discrete labels (e.g., changing a label from relevant to irrelevant) according to task-specific criteria. For a detailed formalisation of this paradigm, readers are referred to \citet{hou2024neuralmachineunranking}. Examples of these tasks are illustrated in Figure~\ref{fig:task.comparison}.}

Removing queries or documents from a neural \gls{ir} model can provoke unintended consequences. As shown in Figure~\ref{fig:task.comparison}, leaving positions vacated by such removals unfilled might lead users to suspect intentional deletion, potentially triggering a `Streisand effect' \citep{chundawat2023can} which refers to the idea that attempts to hide information paradoxically increase attention to it. Conversely, filling the vacated positions with subsequent documents based on relevance ranking could potentially insert less relevant content, which causes performance degradation risks in machine unlearning \citep{Zhang2022,li2024machineunlearningtaxonomymetrics,xu2024machine}. On the other hand, filling these positions with documents ranked by relevance may introduce less relevant content, risking performance degradation in the unlearning process \citep{Zhang2022,li2024machineunlearningtaxonomymetrics,xu2024machine}. Moreover, this strategy may inadvertently expose sensitive information due to positional shifts, creating potential sensitive information leakage risks \citep{xu2024machine,liu2024threatsattacksdefensesmachine}.

Furthermore, some researchers~\citep{warnecke2021machine,goel2024corrective,xu2024machine} argue that machine unlearning should go beyond merely removing specific samples and instead encompass corrections within the training data.
This is crucial for addressing incorrect or polluted data points that require unlearning \citep{cao2018efficient}.
To our knowledge, no existing work has explored data correction in the context of machine unlearning within \gls{ir}.

%\textbf{Motivation of corrective unranking.} 
To mitigate the risks associated with data removal and to integrate error correction capabilities into the neural \gls{ir} models, we propose a novel task termed \emph{neural corrective machine unranking}, or \emph{corrective unranking} for short.
This task incorporates forgetting %\axelin{the task is not meant to forget -- the model is meant to forget}
specific samples previously learned in neural \gls{ir} models while simultaneously learning suitable substitutes to effectuate corrections in the retrieval outputs. State-of-the-art machine unlearning methods, even when combined with finetuning on substitutes, fall short of delivering optimal performance in this task (this observation will be evidenced in Figures~\ref{fig:unlearn.result.correct} and \ref{fig:unlearn.result.retain}). In response, we introduce a novel method specifically designed to handle corrective unranking termed \emph{\gls{curd}}.

Specifically, our contributions are threefold:
\begin{enumerate}
\item We formalise the task of corrective unranking and introduce comprehensive evaluation metrics that assess forgetting, correction, and retention performance.
\item We propose \gls{curd}, a novel teacher--student (i.e., knowledge distillation--based unlearning) framework tailored to corrective unranking. \Gls{curd} induces forgetting by reducing the relevance scores of forgotten query-document pairs to a specified quantile of negative pair scores.
\item We demonstrate that \gls{curd} achieves fast and effective corrective unranking, surpassing state-of-the-art methods in terms of forgetting and correction while maintaining model retention and generalisation capabilities.%\axelin{check that the claim in the last sentence is consistent with the claims in the abstract.}
\end{enumerate}

The rest of the paper is organised as follows: Section~\ref{sec:literature} reviews the related works relevant to our study, providing context and background.  %\axelin{this sentence is too generic. It could be in /any/ paper. Be more specific.}. 
Section~\ref{sec:task} defines the corrective unranking task we aim to address. Section~\ref{sec:method} presents the proposed method for corrective unranking and Section~\ref{sec:exp} describes the experimental setup used to evaluate the method, including datasets, tested neural \gls{ir} models and metrics. Section~\ref{sec:results} discusses the results obtained and evaluates the performance of the proposed method. Finally, Section~\ref{sec:conclusion} concludes the paper.

\section{Related works} \label{sec:literature}

This work focuses on corrective unlearning in information retrieval, which is still largely underexplored. However, existing machine unlearning and corrective unlearning studies could provide insights into this work.

\subsection{General machine unlearning}

Machine unlearning seeks to remove specific data points from a trained model, often motivated by concerns over data privacy~\citep{bourtoule2021machine,xu2023machine} or the presence of poisoned data used in backdoor attacks~\citep{LI2024120135,ZHONG2024120283}. A straightforward way for machine unlearning is retraining models from scratch, but this is often prohibitively costly in terms of computational resources and time~\citep{wu2020deltagrad}. In response, researchers have developed alternative methods.

\subsubsection{Unlearning via influence function and gradient operation}

\emph{Influence function-based methods} adjust model parameters to negate the effect of the to-be-forgotten data \citep{Koh17,Giordano19}. In linear models, methods like Newton's method~\citep{Guo20} reduce data point influence, while projective residual updates~\citep{Izzo21} offer fast parameter adjustments for small groups of data. For strongly convex models, closed-form updates translate data changes into parameter adjustments~\citep{Warnecke23}. Hessian-based Newton methods are resource-intensive, prompting alternatives like the Fisher Information Matrix (FIM)~\citep{Martens_2014} and Neural Tangent Kernel~\citep{Golatkar20Forgetting}.  Expanding on FIM, \citet{Foster_Schoepf_Brintrup_2024} designed a method to induce forgetting by dampening parameters proportionally to their importance relative to the forget set. \jerryRevise{Similarly, \citet{Yang_Han_Wang_Liu_2025} proposed ETR. In its ``Erase'' stage, the FIM is used to identify parameters most critical to the to-be-forgotten samples and their neighboring nodes, followed by targeted modifications that suppress both direct and indirect influence.}

Other common approaches are \emph{gradient-based unlearning} methods, such as Negative Gradient (NegGrad)~\citep{Zhang2022,tarun2023fast} and Amnesiac unlearning~\citep{Graves_Nagisetty_Ganesh_2021} reverse parameter updates tied to specific data batches, while DeltaGrad~\citep{wu2020deltagrad} allows efficient retraining for small data changes. Forsaken~\citep{9844865} employs selective neuron adjustments to control unlearning precision. 
\jerryRevise{More recently, \citet{hoang2024learn} introduced Projected Gradient Unlearning, a method that modifies model parameters by projecting updates onto subspaces orthogonal to gradients important for the retain set. This orthogonal projection ensures that knowledge relevant to the remaining data is preserved. PGU leverages standard Stochastic Gradient Descent (SGD) for weight updates.}

\subsubsection{Unlearning via knowledge distillation}

Another prevalent method is \emph{knowledge distillation-based unlearning}, to which the approach proposed in this work also belongs. 
Knowledge distillation typically involves transferring knowledge from a large, complex teacher model to a smaller, more efficient student model~\citep{wang2021knowledge,gou2021knowledge}.
However, in knowledge distillation-based unlearning, the student model is trained to emulate the teacher model while intentionally omitting the information designated for forgetting. \citet{kim2022efficient} introduced a two-stage unlearning process: the first stage involves training a deep learning model that loses specific information using contrastive labels from the requested dataset; the second stage retrains this model with knowledge distillation to recover its overall performance. To enable data-free unlearning, \citet{10097553} proposed a zero-shot machine unlearning method that utilises error-minimising noise and gated knowledge transfer within a teacher--student learning framework. \citet{chundawat2023can} developed a dual-teacher strategy that utilises misinformation from an incompetent teacher to induce forgetting while a competent teacher helps retain the model's performance. 

Additionally, \citet{NEURIPS2023_062d711f} proposed unbounded unlearning in which the student model selectively retains or discards information based on its alignment with or deviation from the teacher model.  More recently, \jerryRevise{\citet{Zhou_2025_CVPR} introduced a \emph{mask distillation} approach that refines the retention process using “dark knowledge” (i.e., soft targets from non-ground-truth class logits). By applying a mask to separate logits associated with forgetting targets from those representing retained classes, the method jointly optimises both forgetting and retention components.}

As is supported by the empirical evaluation in this paper, both influence functions and gradient-based methods fail to deliver satisfactory unlearning results in the context of neural \gls{ir}.
Additionally, most existing knowledge distillation methods are primarily designed for classification tasks and are incompatible with the neural \gls{ir} setting \cite{hou2024neuralmachineunranking}.

\subsection{Machine unlearning in information retrieval}

To our knowledge, \citet{Zhang2022} were the first to propose a machine unlearning method tailored specifically to the task of image retrieval. However, their approach only considers the removal of to-be-ranked items, neglecting the unlearning of queries. In response, \citet{hou2024neuralmachineunranking} formalised machine unlearning within the broader context of \gls{ir}, 
addressing both query and document (to-be-ranked items); \citet{hou2024neuralmachineunranking} also introduced a model-agnostic unlearning technique for neural \gls{ir}. However, despite its versatility, the performance of their method is sensitive to the choice of certain tuning parameters. 
 
\subsection{Corrective unlearning}

Recent discussions among researchers~\citep{warnecke2021machine,xu2024machine} suggest that machine unlearning should not only focus on removing specific samples but also involve corrections within the training data. However, unlearning methods that incorporate corrections have received relatively little attention. \citet{cao2018efficient} pioneered the concept of \emph{causal unlearning}, which identifies and eliminates the sources of misclassification within the training dataset in an iterative manner, thereby correcting polluted machine learning systems. Recently, \citet{goel2024corrective} have introduced \emph{corrective machine unlearning} to tackle the challenges presented by data corrupted through unknown manipulations in trained models. A key feature of this method is the evaluation of correction accuracy on the manipulated training dataset, which serves to assess the efficacy of the unlearning process in rectifying data corruption.
However, the methodologies proposed by \citet{cao2018efficient} and \citet{goel2024corrective} focus exclusively on classification tasks which are not compatible with neural \gls{ir} systems. This necessitates the development of specialised formalisation and evaluation metrics tailored for the corrective unlearning in neural \gls{ir} contexts.

\jerryRevise{Corrective unlearning bears strong conceptual similarities to knowledge editing in language models. Both aim to adjust a model’s stored knowledge without full retraining. For example, \citet{de-cao-etal-2021-editing} introduce KnowledgeEditor, a hyper-network that predicts weight updates to fix a specific factual “bug” in a language model without retraining. \citet{wang2024knowledge} survey knowledge-based model editing, emphasizing methods that ``precisely modify the large language model (LLM) to incorporate specific knowledge, without negatively influencing other irrelevant knowledge''. Notably, these knowledge editing approaches target text generation tasks: they modify autoregressive or seq2seq LMs (often Transformers) to change outputs of Q\&A or completion prompts.}

\jerryRevise{Knowledge editing or LLM unlearning assumes a Transformer LM with vast implicit knowledge. In the field of \gls{ir}, despite the rise of generative retrieval methods \citep{li2024learning, gienapp2024evaluating} where an LLM directly predicts identifiers or text as retrieval results, the dominant approach remains the \emph{learning-to-rank} architecture \citep{zhao2024dense}. Moreover, \gls{ir} rankers often use non-autoregressive networks (e.g. BERT encoders \cite{hofstatter2021efficiently}, dual encoders\cite{ColBERT2020}, etc.) and produce a ranked list of existing documents. This architectural difference means that techniques developed for editing LLMs do not directly apply to \gls{ir} models. }

\section{Problem definition}  \label{sec:task}
\subsection{Neural information retrieval}
Let $\querySpace$ denote the set of all possible queries and $\documentSpace$ the set of all possible documents. Define $\datasetSpace$ as the universe of possible datasets. A dataset for (neural) \gls{ir}, $\originalSet \in \datasetSpace$, then consists of tuples $(x, y)$, where:
\begin{itemize}
    \item $x = \langle q, d \rangle \in \querySpace \times \documentSpace$ represents a query--document pair;
    \item $y \in \{ \text{+}, \text{-} \}$ is the ground-truth relevance label for $\langle q, d \rangle$, where `+' indicates the document is relevant to the query, and -' indicates the document is irrelevant. Note that ordinal relevance labels such as `highly relevant' or `partially relevant' can always be reduced to binary labels by applying an appropriate threshold.
\end{itemize}
Hereafter, we assume that every query--document pair in the dataset is unique, i.e.\ that each query--document pair $\langle q,d\rangle$ is associated with a unique relevance label $y_{q,d} \in \{ \text{+}, \text{-} \}$ in the dataset;
and we introduce the following notation. The set of all queries contained in the training dataset is
\begin{align}
    Q \coloneqq \smash{\{q' \in \querySpace \mid \exists \, (\langle q, d\rangle, y) \in \originalSet: q' = q\}}.
    \label{eq:querySet}
\end{align}
The set of all documents contained in the training dataset is
\begin{align}
    D \coloneq \bigcup_{q \in Q} D_q,
\end{align}
where, for each $q \in Q$,
\begin{align}
    D_q \coloneqq \smash{\{d \in \documentSpace \mid \exists\, (\langle q', d'\rangle, y) \in \originalSet: \langle q', d'\rangle = \langle q, d\rangle \}}
\end{align}
is the set of documents for which the training dataset contains relevance labels w.r.t.\ Query~$q$, i.e., these are the `to-be-ranked' documents for $q$; and this set can be further partitioned into the subset of \emph{positive} (relevant) documents, $\smash{D_q^+} \coloneqq \smash{\{d \in D_q \mid y_{q,d} = \text{+}\}}$, and \emph{negative} (irrelevant) documents, $\smash{D_q^-} \coloneqq \smash{\{d \in D_q \mid y_{q,d} = \text{-}\}}$, 
With this notation and under the uniqueness assumption mentioned above, we may alternatively express the training dataset as:
\begin{align}
    \originalSet =  \{(\langle q, d\rangle, y_{q,d}) \mid q \in Q, d \in D_q\}.
\end{align}

Let $\parameterSpace \subseteq \reals^p$ be the \emph{parameter space}, i.e.,  $w \in \parameterSpace$ is the collection of parameters (e.g., weights) of a \emph{neural ranking model}. For brevity, we will refer to the elements  $w \in \parameterSpace$ themselves as (neural ranking) models. A neural ranking model $w \in \parameterSpace$ is trained to predict a relevance score $f_w(x) \in \reals$ for a query--document pair $x = \langle \singleQuery, \singleDocument\rangle$. The relevance scores are used to rank documents such that those with higher scores are positioned earlier in the search results for the query. We define a \emph{trained} neural ranking model as:
\begin{align}
  \trainedModel \coloneqq \mathop{\mathrm{arg\,min}}_{w \in \parameterSpace} L_{\originalSet}(w),
\end{align}
where $L_{\originalSet}(w)$ is a suitable loss which penalises the discrepancy between the prediction of Model
~$w$ and the ground truth in $\originalSet$. Note that while our definition of neural \gls{ir} is framed within a pairwise paradigm, training objectives can be pointwise, pairwise, or listwise. For more information on the training objectives of neural ranking models, see \citet{guo2020deep}.

\subsection{Machine unlearning in information retrieval}

\jerryRevise{As previously discussed, machine unlearning in \gls{ir} can be categorised into two forms: document removal and query removal. Both aim to eliminate learned relevance signals for specific query-document pairs, rather than to remove the query or document entities themselves. That is, unlearning in this context operates at the level of relevance judgments, not the underlying data records. To formalise these two paradigms, we define:}

\begin{itemize}
    \item \emph{Document removal}, depicted on the left-hand side of Figure~\ref{fig:task.comparison}, seeks to ensure that documents in some set $D^f \subseteq D$ are no longer considered to be relevant for any query. In this case, we may define a \emph{forget} set as 
    \begin{align}
    \forgetSet \coloneqq \{(\langle \singleQuery, \singleDocument \rangle, y) \in \originalSet \mid \singleDocument \in \documentSet^f \}.
\end{align}
    
    \item \emph{Query removal}, depicted on the right-hand side of Figure~\ref{fig:task.comparison}, seeks to ensure that queries in some set $Q^f \subseteq Q$ no longer have their positive (i.e., relevant) documents retrieved. In this case, the \emph{forget set} is
    \begin{align}
    \forgetSet \coloneqq \{(\langle \singleQuery, \singleDocument \rangle, y) \in \originalSet \mid \singleQuery \in \querySet^f \text{ and }  y=\text{+}\}.
\end{align}
\end{itemize}
Note that query and document removal may occur simultaneously. In this case, the resulting forget set is the union of the individual forget sets.

We may then define a \emph{retain} set as $\retainSet \coloneqq \originalSet \setminus \forgetSet$.
Within this work, a machine unlearning model should replicate the performance of the \emph{retrained} model, i.e., of the model which is retrained from scratch on $\retainSet$, while simultaneously lowering the relevance scores of any query--document pair in the forget set, i.e., for $(\langle \singleQuery, \singleDocument \rangle, y) \in \forgetSet$. This ensures that $d$ is not retrieved in response to $q$. However, as illustrated in Figure~\ref{fig:task.comparison}, simply omitting certain retrievals—whether by leaving top-ranked positions empty or by skipping to lower-ranked documents—can introduce issues such as the `Streisand effect' or misinformation.

To address these concerns, we propose an alternative approach: substituting to-be-forgotten documents with suitable alternative documents.

\subsection{Corrective unranking}

As discussed and illustrated in Figure~\ref{fig:task.comparison}, directly removing retrieved items is not a robust, necessitating the substitution of documents at the positions from which documents are removed, whether for query or document removal. We term this approach \emph{neural corrective machine unranking}, or simply \emph{corrective unranking}.

Both document removal and query removal involve ensuring that certain documents are not retrieved in response to certain queries. In a list-wise manner, we define the set of documents to be corrected for a query $q$ as:
\begin{align}
\documentSet_q^f \coloneqq \{d \in D_q \mid \exists \, (\langle q', d' \rangle, y') \in \forgetSet: \langle q, d\rangle = \langle q', d'\rangle\}.
\end{align}
Let $\smash{r_q\colon D_q^f \to \documentSpace \setminus D_q^f}$ be the function mapping each document $d \in \smash{D_q^f}$ to a suitable \emph{substitute document} $r_q(d) \in \smash{\documentSpace \setminus D_q^f}$. To simplify the notation, we also define $r(\langle q, d\rangle) \coloneqq \langle q, r_q(d)\rangle$ to be the `corrected' query--document pair which results from replacing the document $d$ in the query--document pair $\langle q, d\rangle$ by its substitute document.

With this notation, we can define a \emph{substitute} set as:
\begin{align}
  \forgetSet^* 
  & \coloneqq \{(\langle \singleQuery, r_\singleQuery(\singleDocument) \rangle, y) \mid (\langle \singleQuery, \singleDocument \rangle, y) \in \forgetSet\}
  = \{(r(x), y) \mid (x, y) \in \forgetSet\}.
\end{align}
%\axel{introducing the notation $r(x)$ greatly simplifies the notation in the following sections, I think.}
The objective of corrective unranking is then formulated as:
\begin{align}
  \correctiveModel \coloneqq \mathop{\mathrm{arg\,min}}_{w \in \parameterSpace} L_{\originalSet^*}(w),
   \label{eq:corrective.objective}
\end{align}
where $w$ is initialised from $\trainedModel$ and $\originalSet^* \coloneqq \forgetSet^* \cup \retainSet$ is the modified training dataset in which all to-be-forgotten documents have been replaced with suitable substitutes, unless they were already considered irrelevant (negative) for a particular query. Note that the relevance labels remain unchanged.

\section{Proposed method for corrective unranking} \label{sec:method}

\glsreset{curd}

\subsection{Overview}

\jerryRevise{The goal of corrective unranking, i.e., Objective~\eqref{eq:corrective.objective}, is to modify a pre-trained ranking model such that certain query-document pairs are effectively ``forgotten'' while preserving the model’s general utility and, where necessary, correcting the ranking by integrating substitute information.}

\jerryRevise{Retraining a model from scratch on a modified dataset is often impractical due to high computational costs and potential performance instability  \cite{NEURIPS2024_16e18fa3}. To address this challenge, we adopt a teacher–student distillation strategy, which is commonly used in machine unlearning \citep{chundawat2023can,NEURIPS2023_062d711f}. In this framework, the original model  $\trainedModel$ acts as a teacher, providing target outputs (i.e., relevance scores) for the student model $w$, which is updated to satisfy new constraints introduced by forgetting, correcting, and retaining requirements.
}

\begin{figure}[t]
\centering
\includegraphics[width=0.7\textwidth]{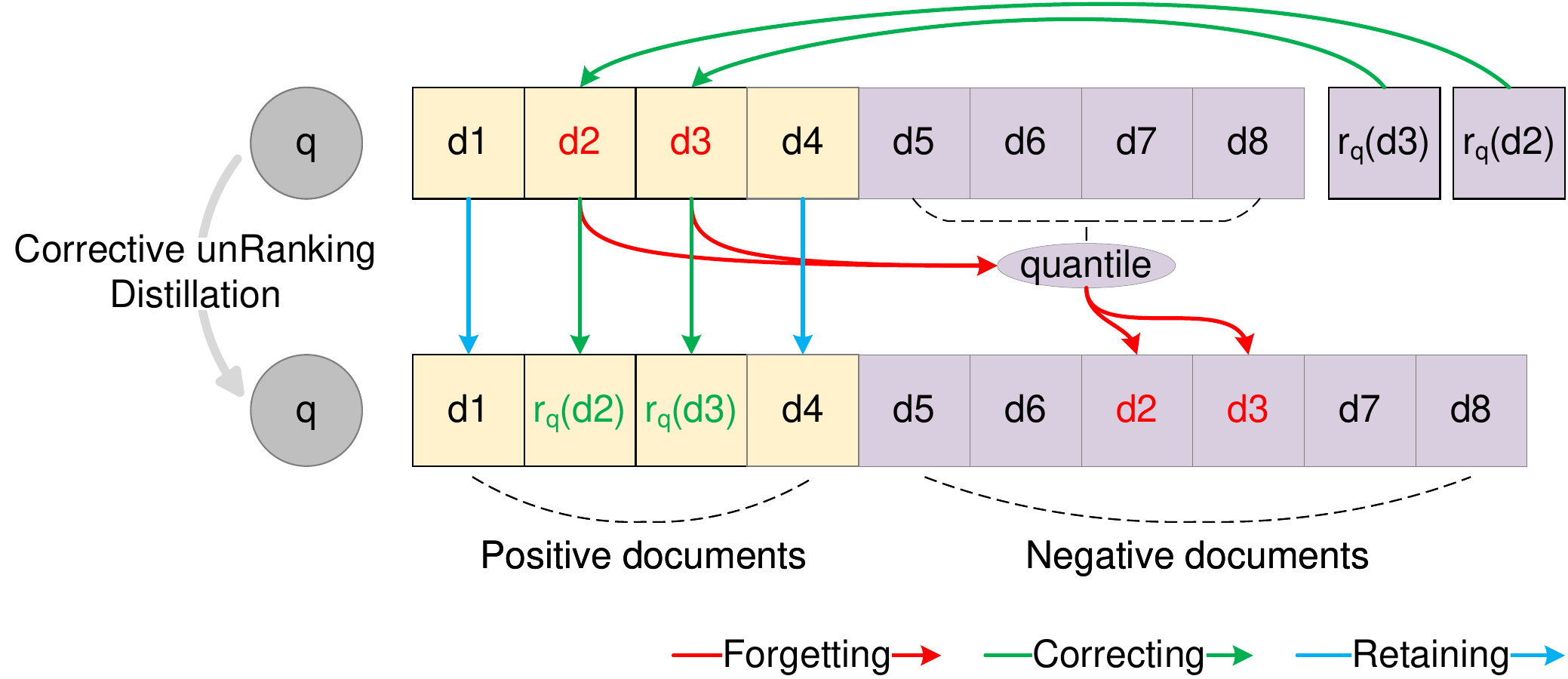} 

\caption{An illustration of the \gls{curd} strategy. Consider the case where documents `d2' and `d3' are to be forgotten for a given query `q'. The \gls{curd} strategy adjusts the relevance scores of the substitute documents `$\mathrm{r_d(d2)}$' and `$\mathrm{r_d(d3)}$' to closely match those of `d2' and `d3', respectively. Additionally, \gls{curd} modifies the relevance scores of `d2' and `d3' to align with those of negative documents for `q'. Furthermore, \gls{curd} ensures that the relevance scores of the remaining documents are preserved.}
\label{fig:forgetting_correcting_demo}
\end{figure}

\begin{enumerate}
\item \textit{Forgetting.} 
For each $(\langle q, d\rangle, y) \in \forgetSet$ we wish to reduce the relevance score produced by $w$ far enough such that $d$ is no longer retrieved in response to Query~$q$. To achieve this, our updates seek to train $w$ to satisfy
\begin{align}
  f_w(\langle q, d\rangle) \approx \quantile_{\trainedModel}^\gamma(q; D_q^-),    \label{eq:forgetting.overview}
\end{align}
for any $(\langle q, d\rangle, y) \in \forgetSet$, where
\begin{align}
  \quantile_{\mathcal{M}}^\gamma(q; D_q^-) \coloneqq \mathrm{quantile}^\gamma(\{f_{\mathcal{M}}(\langle q, d^-\rangle) \mid d^- \in D_q^-\}),
  \label{eq:forgetting.quantile}
\end{align}
denotes the $\gamma$-quantile ($\gamma \in [0,1]$) of the relevance scores associated with the full set of negative documents for Query~$q$ under model $\mathcal{M}$.

\item \textit{Correcting.} For each to-be-forgotten document, we wish to introduce the substitute document into the vacated position in the retrieved list to correct the ranking. That is, for each $(x, y) \in \forgetSet$, our updates seek to train $w$ to satisfy 

\begin{align}
   f_{w}(r(x)) \approx f_{\trainedModel}(x).
   \label{eq:correcting.overview}
\end{align}
    
\item \textit{Retaining.} The relevance scores of all query--document pairs in the retain set should remain roughly unchanged. Therefore, for any $(x, y) \in \retainSet$, our updates seek to train $w$ to satisfy 
    \begin{align}
        f_{w}(x) \approx f_{\trainedModel}(x).
        \label{eq:retain.overview}
    \end{align}
\end{enumerate}
We refer to this distillation strategy as \emph{\gls{curd}}. Figure~\ref{fig:forgetting_correcting_demo} illustrates this approach. \jerryRevise{To operationalise this strategy, \gls{curd} introduces three loss components, each designed to address a specific objective: forgetting unwanted associations, correcting rankings by incorporating substitutes, and retaining knowledge. These components are optimised jointly using gradient-based updates, forming a unified objective.}

\subsection{Corrective unranking distillation}

We now describe how \gls{curd} can be implemented in practice. We employ a teacher-student framework, where $ \mathcal{M} = \trainedModel $  serves as the teacher model and $ w $ as the student model. This implementation comprises two components: one for concurrently performing forgetting and correction (which seeks to ensure \eqref{eq:forgetting.overview} and \eqref{eq:correcting.overview}, and the other dedicated to retention (which seeks to ensure \eqref{eq:retain.overview}).

\paragraph{Forgetting and correcting}
This component is designed to decrease the relevance scores for `to-be-forgotten' documents while simultaneously increasing the scores for their substitutes. 

To approximately achieve \eqref{eq:forgetting.overview} and \eqref{eq:correcting.overview},  we define the forgetting and correcting loss in \gls{curd} as:
\begin{align}
    \loss_{\mathcal{M}}^{\text{fc}}(w) 
    & \coloneqq \smashoperator[lr]{\sum_{(x, y) = (\langle q, d\rangle, y) \in \forgetSet}}
    \hingeLoss(f_w(x), \quantile_{\mathcal{M}}^\gamma(q; D_q^-))  + 
    \hingeLoss(f_{\mathcal{M}}(x), f_w(r(x))), \label{eq:correction}
\end{align}
where $\hingeLoss(a, b) = \max\{0, a-b\}$ denotes the \emph{hinge loss.} This choice of loss is motivated by the observation that $f_w(x)$ is typically initially larger than $ \quantile_{\mathcal{M}}^\gamma(q; D_q^-)$, while $f_w(r(x))$ is smaller than $f_{\mathcal{M}}(x)$.

To reduce the computational complexity of computing the quantiles in \eqref{eq:correction}, we replace $\quantile_{\mathcal{M}}^\gamma(q; D_q^-)$ in \eqref{eq:correction} by  $\quantile_{\mathcal{M}}^\gamma(q; A_q^-)$, where $\smash{A_q^-}$ is a subset of $\smash{\documentSet_q^-}$ of size $k$ (e.g., $k = 5$ or $k = 10$) which is sampled once at the start of the unlearning algorithm. 

\paragraph{Retaining}
To ensure that the relevance scores produced by $w$ closely align with those generated by $\mathcal{M}$ across all samples retained as per~\eqref{eq:retain.overview}, we introduce the \emph{retaining loss} to minimise the discrepancy in scores for both positive and negative documents.

The retaining loss is then defined as: 
\begin{align}
    \loss_{\mathcal{M}}^{\text{r}}(w) 
    & \coloneqq \sum_{q \in Q} \sum_{d_q^+ \in D_q^+ \setminus D_q^f} \biggl[\hingeLoss \bigl(f_\mathcal{M}(\langle q, d_q^+ \rangle),f_w(\langle q, d_q^+ \rangle) \bigr) \nonumber\\
    & \quad + \frac{1}{|A_q^-|}\smashoperator[r]{\sum_{d_q^- \in A_q^-}}\hingeLoss\big(f_w(\langle q, d_q^- \rangle),f_\mathcal{M}(\langle q, d_q^- \rangle)\big) \biggr],
    \label{eq:retain}
\end{align}
where for each $d_q^+ \in D_q^+ \setminus D_q^f$, we $A_q^-$ is again a subset of $\smash{\documentSet_q^-}$ of size $k$ sampled once at the start of the unlearning algorithm. 

\jerryRevise{In practice, we recommend setting the size of $A_q^-$ to 5, as this yields a favorable balance between retention performance and computational efficiency. Larger values (e.g., 10 or 15) may provide marginal improvements in retention performance but incur greater computational cost, as demonstrated in our sensitivity analysis (Section~\ref{chap:sensitivity}). For the factor $\gamma$, a default value of 0.0 is suggested to emphasize forgetting and correction fidelity. However, in scenarios where retention accuracy is more critical, higher values of $\gamma$  (e.g., 0.25 or 0.50) may be more appropriate.}

\paragraph{Overall corrective unranking loss}
The overall corrective unranking objective combines both the forgetting and correcting and retaining components:
\begin{align}
    \mathcal{M}_\mathrm{correct} & \coloneqq \mathop{\mathrm{arg\,min}}_{w} \lambda^{\text{fc}}\loss_{\mathcal{M}}^{\text{fc}}(w) + \lambda^{\text{r}}\loss_{\mathcal{M}}^{\text{r}}(w),
\end{align}
\jerryRevise{In practice, we simply fix $\lambda^{\text{fc}} = 1$ and $\lambda^{\text{r}} = 1$ to assign equal weight to the forgetting–correction and retention objectives. These coefficients are not treated as tunable hyperparameters in this study, but are instead introduced to generalise the corrective unranking loss formulation. Depending on the application context, users may wish to prioritise either correction or retention, thereby enabling potential adaptation of CuRD to domain-specific requirements.}

\jerryRevise{The loss functions defined within \gls{curd} are convex with respect to the relevance scores and can be optimised using standard gradient-based training procedures. Under mild conditions—such as bounded gradients and properly decayed learning rates—the training process is guaranteed to converge to a local optimum. Additionally, since the proposed updates are targeted and localised to the forget and correction sets, \gls{curd} exhibits stability under small perturbations to the training data. This robustness makes it well-suited for deployment in dynamic environments. Empirical support for these convergence and stability properties is provided in Section~\ref{chap:sensitivity}.}

We recall that $\mathcal{M} = \trainedModel$ is the trained model. For a detailed implementation of \gls{curd} in the pair-wise training dataset setting, refer to Algorithm~\ref{algo:curd}.

\begin{algorithm}[htbp]
\caption{\glsdesc{curd}}
\begin{algorithmic}[1]
\Require Trained model $\mathcal{M}$, forget set $\forgetSet$, retain set $\retainSet$, probability $\gamma \in [0,1]$, 
number of epochs $n$, number of negative documents $k$, $\lambda^{\text{fc}}, \lambda^{\text{r}} > 0$
\ForAll{$q \in Q$} \Comment{For each query $q$ in the training set [see \eqref{eq:querySet}].}

    \If{$k \leq \lvert D_q^-\rvert$}
        \State Sample $k$ documents from $D_q^-$ without replacement to obtain $A_q^-$
    \Else{}
        \State Sample $k$ documents from $D_q^-$ with replacement to obtain $A_q^-$
    \EndIf
\EndFor
\ForAll{$(\langle q, d \rangle, y) \in \originalSet = \forgetSet \cup \retainSet$}

    \If{$(\langle q, d \rangle, y) \in \forgetSet$}
        \State ${\quantile} \gets \quantile_{\mathcal{M}}^\gamma(q; A_q^{-})$
    \Comment{ 
    Quantile defined in \eqref{eq:forgetting.quantile}.}
        \State $d^* \gets r_q(d)$ \Comment{Substitute document for $d$.} 
        \State Append $(\langle q, d \rangle, \langle q, d^* \rangle, \quantile)$ to $X$; and $0$ to $Y$ 
        \ElsIf{$y == \text{+}$}
        \State Append $(q, d,A_q^{-})$ to $X$; and $1$ to $Y$
    \EndIf
\EndFor

\State Initialise $w = \mathcal{M}$
\For{$\text{epoch} = 1 \to n$}
    \For{$i = 0 \to |Y| - 1$}
        \State $l \gets 0$
        \If{$Y[i] == 0$}
            \State $l \gets l + \lambda^{\text{fc}}\hingeLoss \big(f_w(X[i][0]),X[i][2])\big)$ \Comment{Forgetting loss in \eqref{eq:correction}.}
            \State $l \gets l + \lambda^{\text{fc}}\hingeLoss \big(f_{\mathcal{M}}(X[i][0]), f_w(X[i][1])\big)$ \Comment{Correction loss in \eqref{eq:correction}.}
        \Else
            \State $l \gets \lambda^{\text{r}} \hingeLoss \big(f_{\mathcal{M}}(\langle X[i][0],  X[i][1] \rangle), f_w(\langle X[i][0],  X[i][1]\rangle)\bigr)$
            \Comment{Retain loss on positive pairs in \eqref{eq:retain}.}
            \ForAll{$d^- \in X[i][2]$} 
                \State $l \gets l + \frac{\lambda^{\text{r}}}{k} \hingeLoss\big(f_w(\langle X[i][0], d^- \rangle),f_\mathcal{M}(\langle X[i][0], d^- \rangle) \big)$
                \Comment{Retain loss on negative pairs in \eqref{eq:retain}.}
            \EndFor
        \EndIf
        \State Update $w$ based on $l$
    \EndFor
\EndFor
\end{algorithmic}
\label{algo:curd}
\end{algorithm}

\section{Experimental Setup} \label{sec:exp}

\subsection{Datasets} We use two \gls{ir} datasets: MS MARCO~\citep{craswell2021ms} and TREC CAR~\citep{dietz2019trec}. To adapt these datasets for corrective unranking, we made the following adjustments:
\begin{itemize}
    \item For each query (in both the training and test datasets), the ratio of positive to negative documents was set to approximately 1:100.
    
   \item In the training dataset, we defined four forget sets based on the fraction of to-be-forgotten query--document pairs among all query-positive document pairs: \emph{Forget-1\,\%}, \emph{Forget-5\,\%}, \emph{Forget-10\,\%}, and \emph{Forget-20\,\%}.
   
    \item The samples in all four forget sets were randomly selected. Within each forget set, we randomly chose some queries for query removal and selected documents (each relevant to at least one query) for document removal. Within each forget set, the number of to-be-forgotten pairs for query removal and document removal was roughly balanced. The remaining pairs were used as the retain set.
    
    \item For each to-be-forgotten query--document pair in the forget set, a substitute document was randomly assigned from the document set, excluding the positive documents for that query.
\end{itemize}

\subsection{Neural ranking models}
Neural ranking models based on \glspl{plm} have demonstrated superior retrieval performance compared to pre-BERT \gls{nir} approaches~\citep{zhao2024dense}. Therefore, this work focuses on \glspl{plm}-based neural ranking models. Four models were selected for evaluation:
\begin{itemize}
    \item \glsdesc{bertcat} (\gls{bertcat})~\citep{hofstatter2021efficiently}.
    \item \glsdesc{bertdot} (\gls{bertdot})~\citep{hofstatter2021efficiently}.
    \item \glsdesc{colbert} (\gls{colbert})~\citep{ColBERT2020}.
    \item \glsdesc{parade} (\gls{parade}) \citep{li2023parade}.
\end{itemize}
Among these models, \gls{bertcat} employs a cross-encoder architecture, where a query and a document are concatenated and separated by the ``[SEP]'' token. The remaining three models use a bi-encoder architecture, in which queries and documents are encoded separately.

\subsection{Evaluation metrics}

The task of corrective unranking extends neural machine unranking by requiring the provision of suitable substitutes during data removal. To evaluate corrective unranking models, we focus on three key aspects: correcting, forgetting, and retaining (on both training and test datasets). In what follows, for some set of documents $D$ and some given model $w \in \parameterSpace$, we let $\rank_{w}(q, d; D)$ 
% \begin{align}
%     \rank_{w}(x; D) = \rank_{w}(q, d; D),
% \end{align}
denote the rank of Document~$d \in D$ among all the documents in $D$ as predicted by the model $w$.

\subsubsection{Correction performance metric} 
To measure the correction performance, we use the following score which is large if the squared difference in reciprocal ranks of the to-be-forgotten and substitute documents is small:
    \begin{align}
     \!\mathbf{P}_{\text{correct}}
     & \coloneqq 
     1- \frac{1}{\lvert \forgetSet \rvert} \sum_{(\langle q, d \rangle, y)) \in \forgetSet}
    \Big( 
    \frac{1}{\rank_{\trainedModel} (q, d; D_q)} -\frac{1}{\rank_{w}(q, d^*; D_q^*)}
    \Big)^{\mathrlap{2}}.\!
    \end{align}
Here, 
\begin{align}
    D_q^* 
    & \coloneqq \smash{\{d \in \documentSpace \mid \exists\, (\langle q', d'\rangle, y) \in \originalSet^*: \langle q', d'\rangle = \langle q, d\rangle \}}\nonumber \\
    & =  \smash{(D_q \setminus D_q^f) \cup \{r_q(d) \mid d \in D_q^f\}},
    \label{eq:substitute_docs}
\end{align}
% $\smash{D_q^*} \coloneqq \smash{(D_q \setminus D_q^f) \cup \{r_q(d) \mid d \in D_q^f\}}$ 
is the set of to-be-ranked documents for Query~$q$ in the modified training dataset $\originalSet^*$.

\subsubsection{Forgetting performance metric} To measure the forgetting performance, we use the mean reciprocal rank of the first document retrieved by $w$ that is designated for forgetting:
\begin{align}
    \mathbf{P}_\text{forget}
    & \coloneqq \frac{1}{\lvert Q_\forgetSet \rvert} \sum_{q \in Q_\forgetSet} \frac{1}{\min\{\mathrm{rank}_{w}(q, d; D_q) \mid d \in D_q^f\}},
\end{align}
where $Q_\forgetSet \coloneqq \{q \in Q \mid D_q^f \neq \emptyset\}$ is the set of queries that appear in the forget set, i.e., the subset of queries from the training dataset that are associated with some document that we wish to substitute.

\subsubsection{Retaining performance metric}  To assess the `retain' and test-set retrieval performance, we use \emph{\gls{mrr}} \citep{liu2009learning} type scores. 
That is, recalling that $\originalSet = \{(\langle q, d\rangle, y_{q,d}) \mid q \in Q, d \in D_q\}$ is the training dataset and letting $\originalSet' = \{(\langle q, d\rangle, y_{q,d}') \mid q \in Q', d \in D_q'\}$ be a test dataset, we define the scores:
\begin{align}
  \mathbf{P}_\text{retain} 
  & \coloneqq 
  \frac{1}{\lvert Q \rvert} \sum_{q \in Q} \frac{1}{\min\{\mathrm{rank}_{w}(q, d; D_q) \mid d \in D_q^+\}},\\ \mathbf{P}_\text{test} 
  & \coloneqq \frac{1}{\lvert Q' \rvert} \sum_{q \in Q'} \frac{1}{\min\{\mathrm{rank}_{w}(q, d; D_q') \mid d \in (D_q')^+\}}.\!\!\!
\end{align}
Here, $\smash{(D_q')^+} \coloneqq \smash{\{d \in D_q' \mid y_{q,d}' = \text{+}\} \subseteq D_q'}$ contains the positive documents for Query~$q$ from the test dataset.

\subsection{Baselines}
Since corrective unranking is a novel task, there are no existing methods directly applicable to it.  Therefore, we have adapted several typical machine unlearning methods to serve as baselines for our experiments:
\begin{itemize}
    \item \textit{Retrain}: Retrain from scratch solely on $\originalSet^*$, i.e., the modified retain set.
    
    \item \textit{Catastrophic forgetting (CF)}: Continue training the model $\trainedModel$ solely on $\originalSet^*$ to induce catastrophic forgetting.
    
    \item \textit{Amnesiac}: Continue training the model $\trainedModel$ with adversarial or noisy data to compel it to forget specific samples~\citep{Graves_Nagisetty_Ganesh_2021}. In this work, Amnesiac is implemented by swapping the rankings of to-be-forgotten documents ($\documentSet_q^f$) and negative documents ($\smash{\documentSet_q^-}$). Subsequently, the optimisation process targets all substitute documents, ensuring they attain relevance scores higher than those of their corresponding negative documents.
    
    \item \textit{Negative Gradient (NegGrad)}: Update the learned model in the reverse direction of the original gradient when unlearning on the forget set, and then fine-tune the updated model with the expanded substitute set. \citep{Zhang2022,tarun2023fast}.
    \item \textit{Bad Teacher (BadT)}: Use a randomly initialised model as a bad teacher on the to-be-forgotten samples, while using $\trainedModel$ as a competent teacher on $\originalSet^*$ \citep{chundawat2023can}.
    
    \item \textit{Selective Synaptic Dampening (SSD)}: A direct model parameter editing method based on the diverse importance of to-be-forgotten and retained samples on $\trainedModel$ \citep{Foster_Schoepf_Brintrup_2024}. We first apply SSD for forgetting and then optimise the model on substitute documents (similar to Amnesaic).
    
    \item \textit{\gls{cocol}}: A knowledge-distillation based machine unlearning method tailored to neural \gls{ir} \citep{hou2024neuralmachineunranking}.  We alternate between using \gls{cocol} and optimise the model on substitute documents (similar to Amnesaic). to implement forgetting and correction, respectively.
\end{itemize}

\subsection{Tuning parameters}
For the Amnesiac and NegGrad methods, we used ten negative documents (10\,\% of all). In \gls{cocol}, the rate and bias coefficients $\alpha$ and $\beta$ were set to 0.5 and 0.0, respectively. 

For \gls{curd}, we set $\smash{k = \lvert A_q^-\rvert}$ to 5  for \gls{colbert} and \gls{bertdot}, and to 10 for \gls{bertcat} and \gls{parade}. The quantile level $\gamma$ in equation \eqref{eq:correction} is set to zero (minimal value).

\section{Results and discussion} \label{sec:results}
\subsection{Corrective unranking performance evaluation} We first fine-tuned each neural ranking model on each dataset, ensuring convergence, i.e., training the models until their performance stabilised, signifying that additional training iterations were unlikely to produce further enhancements. Subsequently, corrective unranking was implemented using \gls{curd} along with seven baseline methods. For performance evaluation, we specifically utilised the \emph{Forget-10\,\%} configuration. We recall that the number of to-be-forgotten pairs designated for query removal and document removal within this set is approximately balanced.

\subsubsection{Performance comparison}

\begin{figure}[ht]
\centering
    \begin{subfigure}{.6\textwidth}
        \centering        \includegraphics[width=\textwidth]{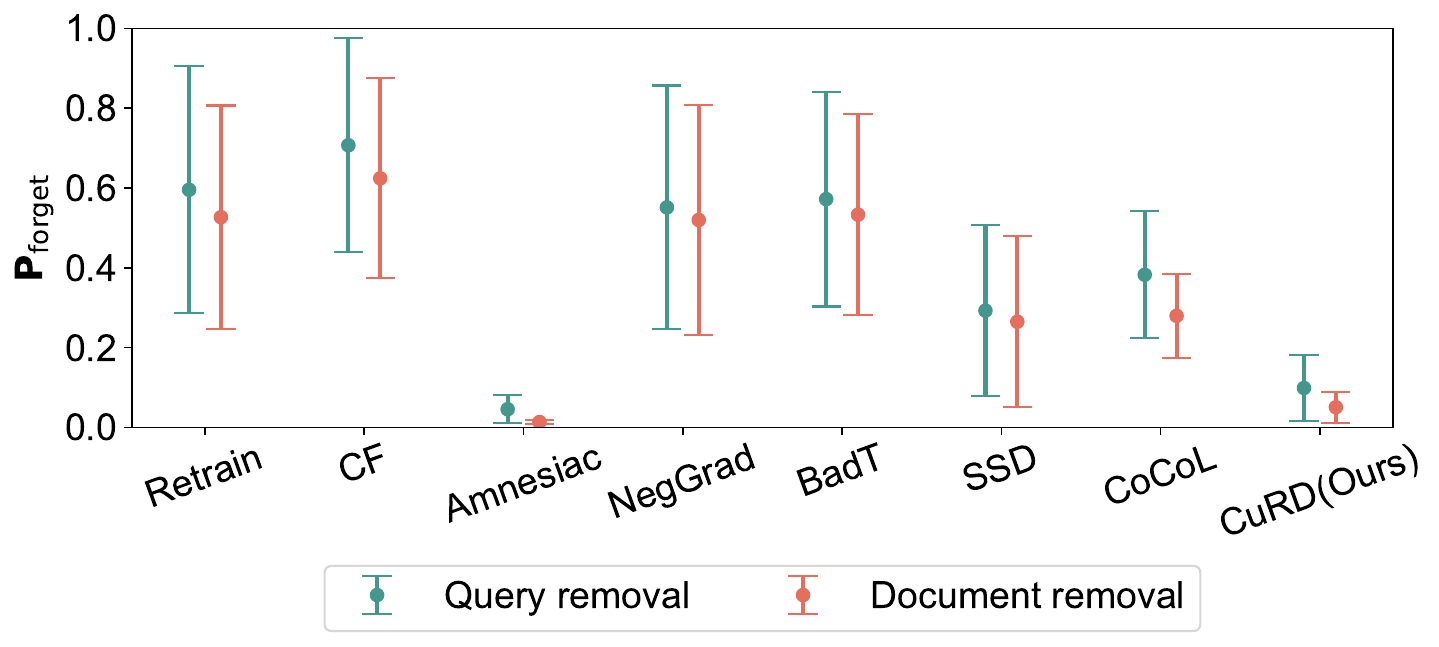} 
\caption{Forgetting performance ($\mathbf{P}_\text{forget}$). Lower values indicate better performance.}
    \label{fig:forget.query}
    \end{subfigure}
    \begin{subfigure}{.6\textwidth}
        \centering
\includegraphics[width=\textwidth]{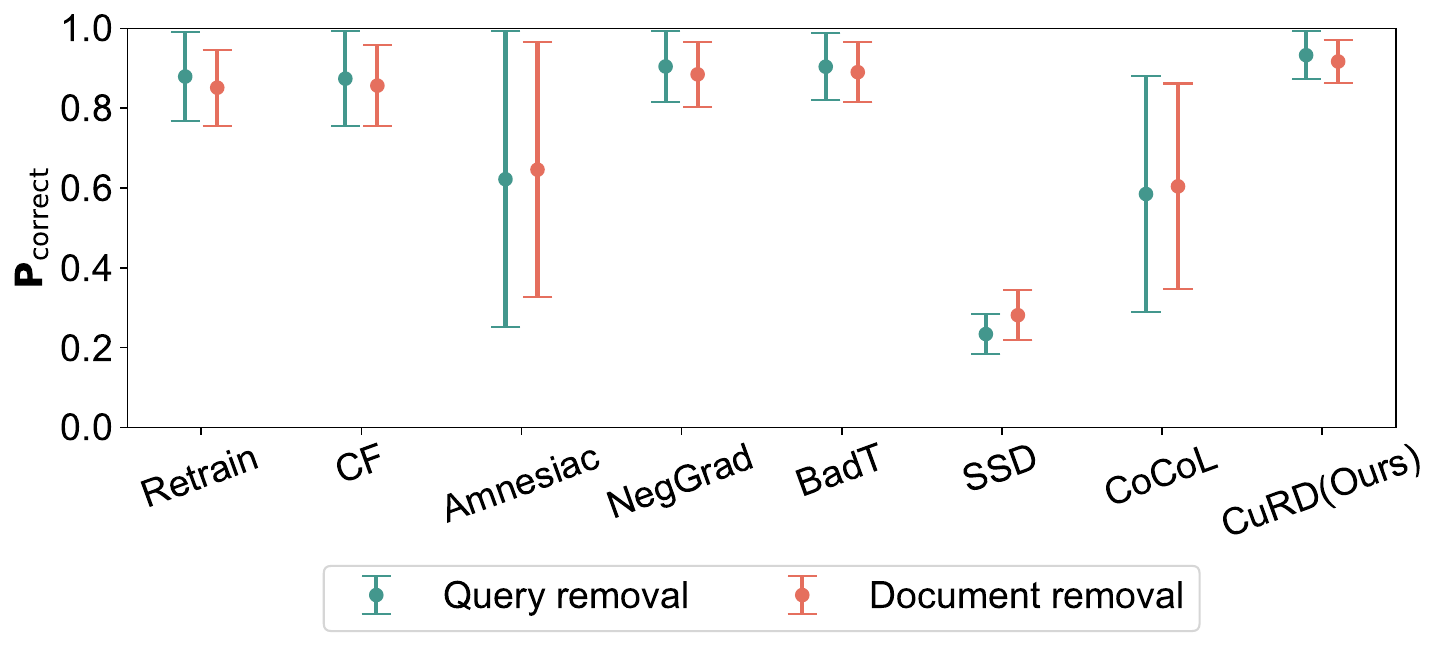} 
\caption{Correction accuracy ($\mathbf{P}_\text{correct}$). Higher values indicate better performance.}
    \label{fig:forget.doc}
    \end{subfigure}
\caption{Correcting unranking performance on forget and substitute sets using \emph{Forget-10\,\%}. The upper and lower bounds of each error bar denote the maximum and minimum performance scores across four neural ranking models on two datasets. This figure illustrates that the proposed method can significantly lower the performance of the to-be-forgotten samples while providing effective corrections of these samples. }
\label{fig:unlearn.result.correct}
\end{figure}

\begin{figure}[ht]
\centering
    \begin{subfigure}{.6\textwidth}
        \centering
\includegraphics[width=\textwidth]{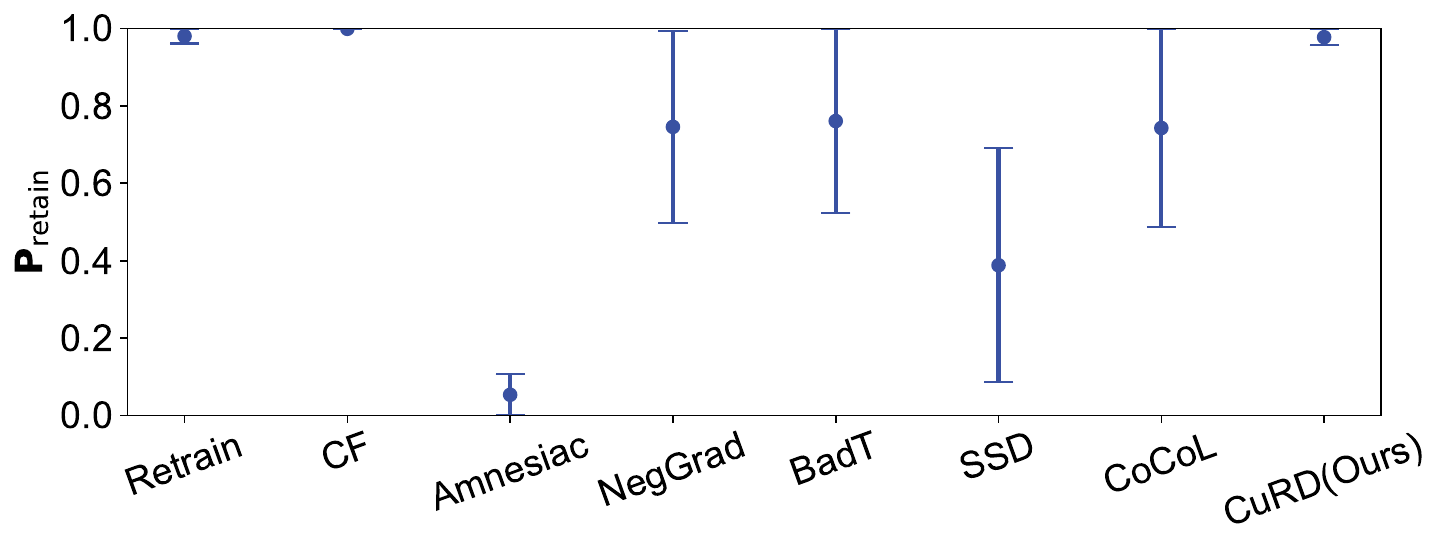} 
    \caption{Model retention performance ($\mathbf{P}_\text{retain}$). Higher values indicate better performance.}
    \label{fig:retain}
    \end{subfigure}
    \begin{subfigure}{.6\textwidth}
        \centering
\includegraphics[width=\textwidth]{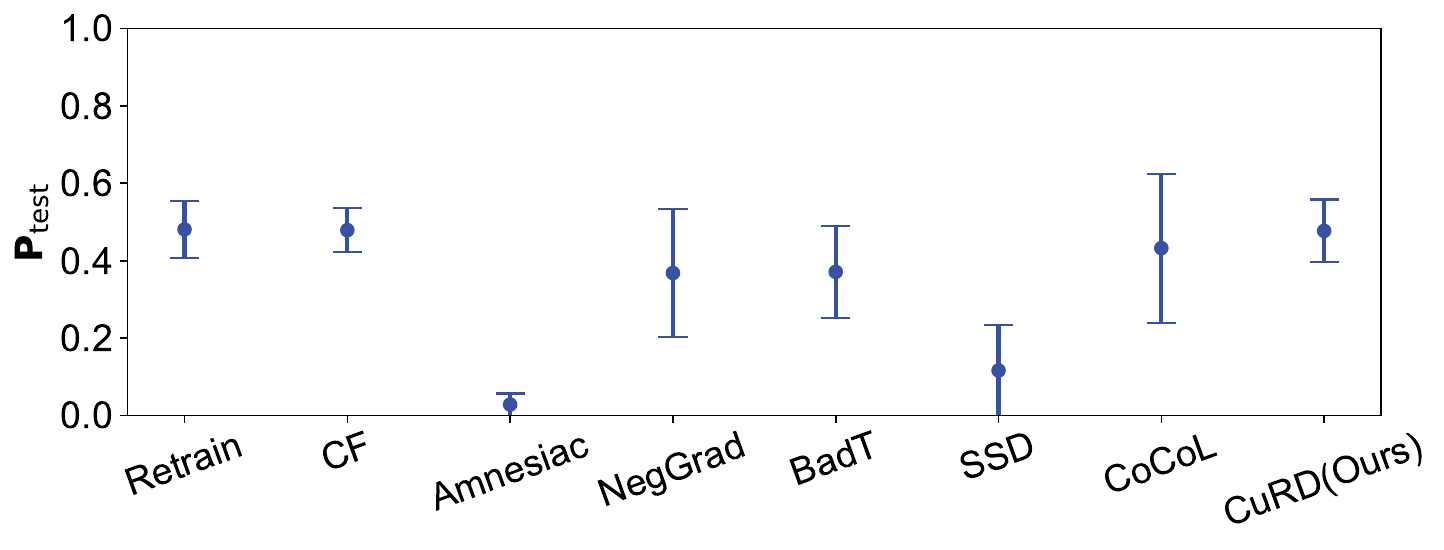} 
    \caption{Model inference  performance ($\mathbf{P}_\text{test}$). Higher values indicate better performance.}
    \label{fig:test}
    \end{subfigure}
\caption{Corrective unranking performance on retain and test datasets. This figure illustrates that the proposed method, \gls{curd}, performs comparably to retraining from scratch on the retain and test datasets, demonstrating that \gls{curd} does not compromise model retention or inference ability. }.

\label{fig:unlearn.result.retain}
\end{figure}

The experimental results are shown in Figure~\ref{fig:unlearn.result.correct} and Figure~\ref{fig:unlearn.result.retain}. For clarity, we visualised the maximum-minimum intervals for each method across the four neural ranking models on both datasets. Detailed results are provided in the Appendix.

\textbf{Forgetting and correcting performance.} Figures~\ref{fig:forget.query} and \ref{fig:forget.doc} illustrate the forgetting and correction performance of various corrective unranking methods on query and document removal tasks, respectively. Among all methods, Amnesiac achieved the most significant forgetting performance, reflected by the lowest performance scores (i.e., $\mathbf{P}_{\text{forget}}$) on the forget set: 0.02 on average for query removal and 0.01 for document removal. \gls{curd} followed with average scores of 0.07 for query removal and 0.04 for document removal. Other methods exhibited larger average forgetting performance scores and broader minimum-maximum ranges. Despite its strong forgetting capability, Amnesiac demonstrated poor correction performance, with high variability and average correction accuracy (i.e., $\mathbf{P}_\text{correct}$) of 0.73 and 0.74 for query and document removal, respectively.
\jerryRevise{ This instability stems from the fact that adversarial or noise-based unlearning disrupts semantic coherence in neural \gls{ir} models, which are highly sensitive to structured relevance patterns \citep{Zhang2022}. The removal of targeted signals using unstructured noise may inadvertently degrade nearby representations, thereby harming the model’s ability to rank appropriate substitutes correctly \citep{7163042}.}

In contrast, \gls{curd} showed superior correction accuracy across all models and datasets, averaging 0.95 for query removal and 0.93 for document removal, surpassing even the Retrain and CF.

\textbf{Retention and generalisation ability.} Figure~\ref{fig:unlearn.result.retain} displays the performance on the retain set and test dataset. Retrain and CF, being fine-tuned on the retain set, excelled with $\mathbf{P}_\text{retain}$ close to 1. \Gls{curd} achieved similar performance, with an average $\mathbf{P}_\text{retain}$ of 0.98. Other baseline methods exhibited larger performance gaps, with wider spans between their maximum and minimum $\mathbf{P}_\text{retain}$. Figure~\ref{fig:test} evaluates generalisation ability via the performance score on an unseen test dataset. \gls{curd} performed comparably to Retrain (0.47) and CF (0.49), with an average score of 0.46 and a minimum above 0.4, outperforming other baselines.

In summary, while Amnesiac excels in forgetting, it compromises retention and generalisation. Fine-tuning methods like Retrain and CF are effective in retention and generalisation but less so in forgetting. \Gls{curd} balances strong forgetting with robust retention and generalisation, combining the strengths of both approaches.

\subsubsection{Unlearn time comparison}

As per \citet{hou2024neuralmachineunranking}, the \emph{normalised unlearn epoch duration} as the average time per unlearning epoch for a model ($w$) divided by the average time per learning epoch for the same model when initially trained $\trainedModel$. The \emph{total unlearn time} is then determined by multiplying the normalised unlearn epoch duration by the total number of unlearn epochs required.

\begin{figure}[ht]
\centering
\includegraphics[width=0.6\textwidth]{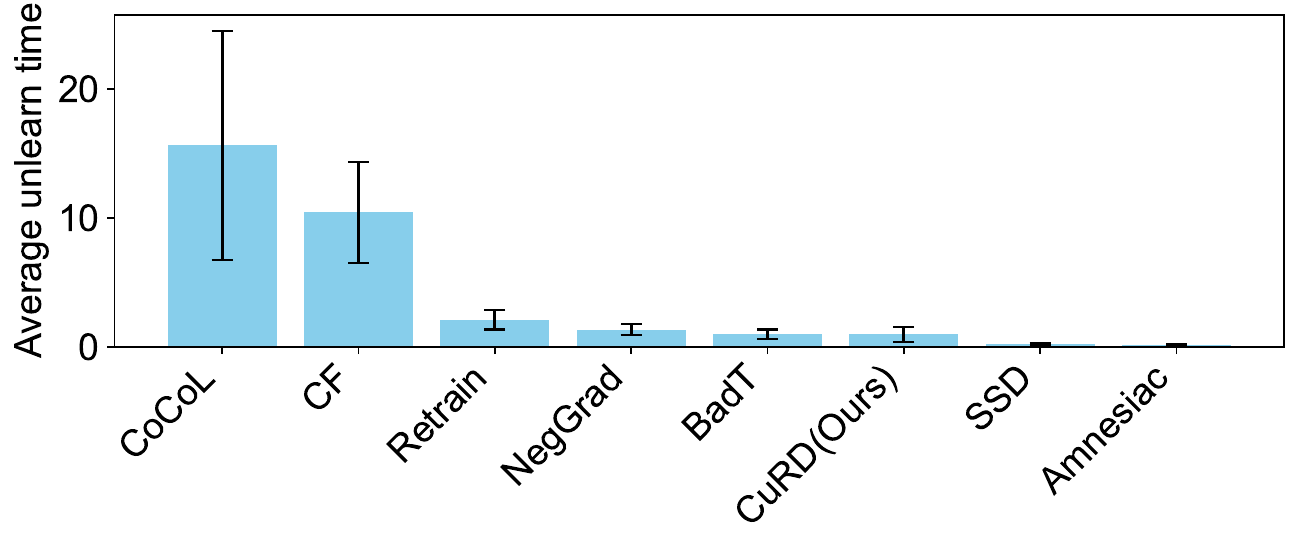} 
\caption{Average unlearning time of \gls{curd} compared to baseline methods. The figure demonstrates that \gls{curd} achieves a shorter unlearning time than most baseline approaches.}

\label{fig:unlearn.time}
\end{figure}

An effective machine unlearning algorithm demands both strong corrective unranking performance and efficient unlearning time. We averaged the unlearning time for all methods across neural ranking models and datasets, as depicted in Figure~\ref{fig:unlearn.time}. The CF and \gls{cocol} methods required the most time, followed by Retrain. In contrast, our method achieved significantly faster unlearning, with an average time of approximately 45.6\,\% that of Retrain. Detailed unlearn time for each experimental group is provided in the Appendix.

\subsection{Parameter sensitivity analysis} \label{chap:sensitivity}

The robustness of \gls{curd} against parameter changes is a key characteristic, primarily influenced by two parameters: the subset size $\smash{A_q^-}$ in Equations~\eqref{eq:correction} and \eqref{eq:retain}, indicating the count of negative documents; and the quantile level $\gamma$ in Equation~\eqref{eq:correction}, which adjusts the relevance score reduction for a to-be-forgotten sample.

We assessed the sensitivity of \gls{curd} by altering the size of $\smash{A_q^-}$ (with values 2, 5, 10, and 15) %\axel{$A_q^-$ is not a `value' but a set. You mean a higher cardinality of this set.} 
and $\gamma$ (at levels 0, 0.25, 0.5, and 0.75, corresponding to minimum, lower quartile, median, and upper quartile). Our results, detailed in Table~\ref{tab:Sensitivity}, demonstrate that \gls{curd} maintains consistent performance across varied settings, as evidenced by mean performance and its standard error over different datasets.

\jerryRevise{As shown in Table~\ref{tab:Sensitivity}, sensitivity experiments reveal that increasing the size of the negative document set $A_q^-$ (i.e., the parameter $k$ in Equations~\eqref{eq:correction} and \eqref{eq:retain}) leads to improved retention performance. This supports the intuition that a richer set of contrastive negatives better preserves knowledge during corrective unranking. \Gls{curd} exhibits robust performance with $k = 5$, and modest gains can be achieved by increasing $k$ when computational resources permit.}

Furthermore, setting $\gamma$ to 0.75 resulted in a higher mean $\mathbf{P}_{\text{forget}}$  for the forget set and a decreased mean $\mathbf{P}_\text{correct}$ for the substitute set, suggesting an optimal upper limit for $\gamma$. These results underline the effectiveness and adaptability of \gls{curd}.

\begin{table}[ht]
\caption{Parameter sensitivity test using mean (performance score)$\pm$(standard error).}
\resizebox{\textwidth}{!}{
\Large
    \begin{tabular}{lrrrrrrr}
        \toprule
          \makecell[l]{Tuning  parameter \\ in \gls{curd}} & \makecell[l]{Parameter \\ value} & {\makecell{$\mathbf{P}_\text{forget}$ \\(query removal)}} & {\makecell{$\mathbf{P}_\text{forget}$ \\ (document \\ removal)}} & {\makecell{ $\mathbf{P}_\text{correct}$ \\(query removal)}} & {\makecell{ $\mathbf{P}_\text{correct}$ \\ (document \\removal)}} & $\makecell{\mathbf{P}_\text{retain}}$ & \makecell{$\mathbf{P}_\text{test}$} \\ \midrule
\multirow{4}{*}{\makecell[l]{Number of \\ negative documents\\ used (the size of $\smash{A_q^-}$ \\ in  Eq~\eqref{eq:correction} and \eqref{eq:retain}) } }   & 2 & 0.056$\pm$0.006 & 0.027$\pm$0.003 & 0.923$\pm$0.010 & 0.904$\pm$0.011 & 0.955$\pm$0.002 & 0.444$\pm$0.003 \\
 & 5 & 0.054$\pm$0.006 & 0.022$\pm$0.002 & 0.964$\pm$0.007 & 0.954$\pm$0.009 & 0.978$\pm$0.001 & 0.450$\pm$0.003 \\
 & 10 & 0.049$\pm$0.005 & 0.023$\pm$0.003 & 0.951$\pm$0.008 & 0.930$\pm$0.010 & 0.990$\pm$0.001 & 0.453$\pm$0.003 \\
 & 15 & 0.054$\pm$0.006 & 0.024$\pm$0.003 & 0.975$\pm$0.007 & 0.963$\pm$0.009 & 0.990$\pm$0.001 & 0.455$\pm$0.003 \\\midrule
\multirow{4}{*}{\makecell[l]{Quantile level\\ ($\gamma$ in Eq~\eqref{eq:correction})} } &  0 & 0.053$\pm$0.006 & 0.024$\pm$0.003 & 0.937$\pm$0.009 & 0.914$\pm$0.010 & 0.986$\pm$0.001 & 0.455$\pm$0.003 \\
 & 0.25 & 0.056$\pm$0.005 & 0.021$\pm$0.002 & 0.922$\pm$0.010 & 0.918$\pm$0.011 & 0.981$\pm$0.001 & 0.439$\pm$0.003 \\
 & 0.5 & 0.059$\pm$0.005 & 0.025$\pm$0.002 & 0.963$\pm$0.008 & 0.953$\pm$0.009 & 0.986$\pm$0.001 & 0.461$\pm$0.003 \\
 & 0.75 & 0.075$\pm$0.006 & 0.041$\pm$0.004 & 0.849$\pm$0.013 & 0.846$\pm$0.013 & 0.984$\pm$0.001 & 0.459$\pm$0.003  \\\bottomrule
\end{tabular}%
}
\label{tab:Sensitivity}
\end{table}

\subsection{Performance of CuRD with varying sizes of forget sets}
\begin{figure}[htbp]
\centering
\includegraphics[width=0.5\columnwidth]{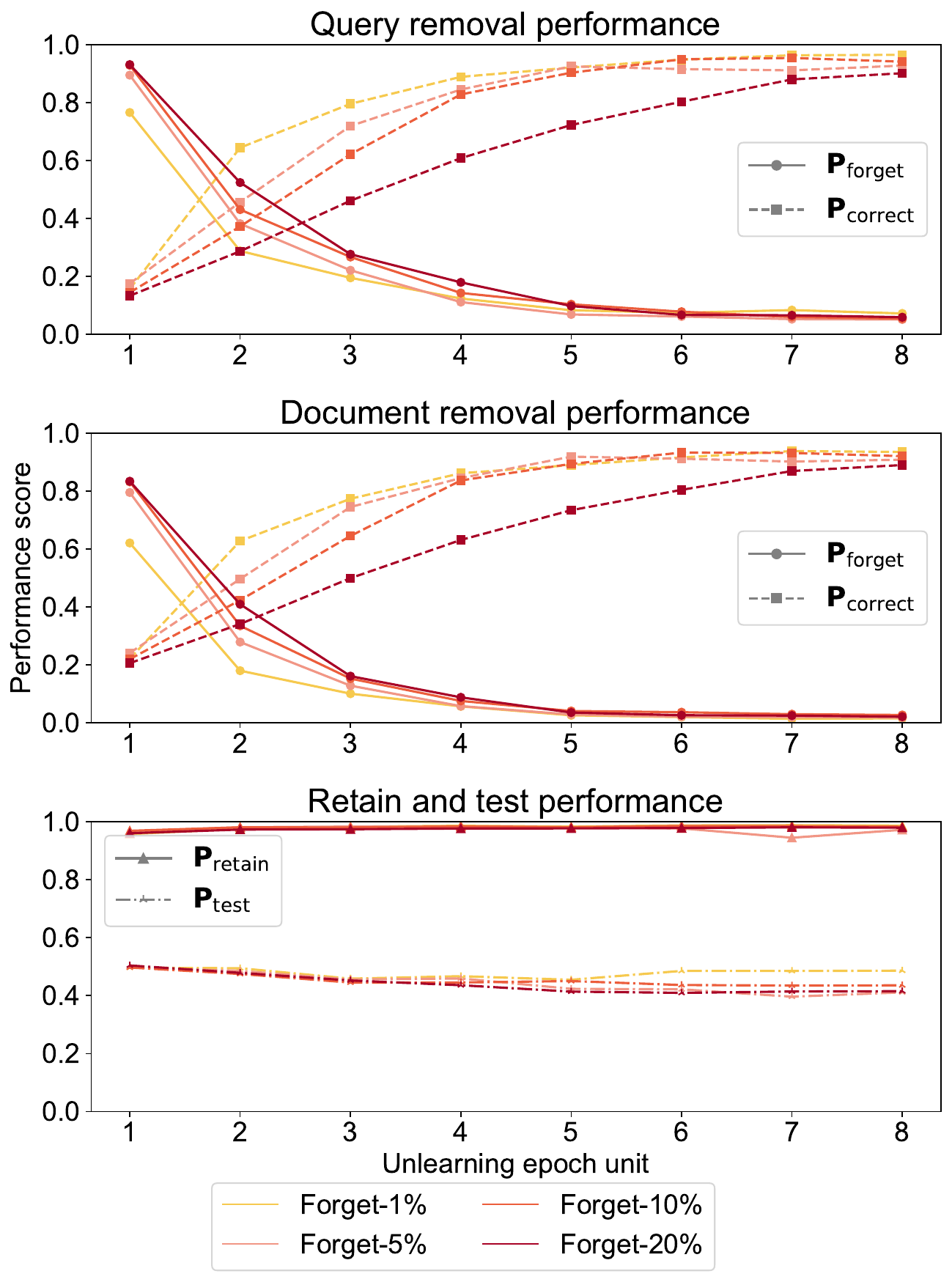}
\caption{Corrective unranking performance with different forget set sizes. To account for variations in convergence speed among different neural ranking models, we introduce the concept of an `epoch unit' for normalised comparison. For \gls{bertcat} and \gls{parade}, one epoch unit corresponds to four epochs, while for \gls{bertdot} and \gls{colbert}, it corresponds to two epochs.  The figure shows that \gls{curd} performs consistently across different forget set sizes.} 

\label{fig:forgetset.volume.exp}
\end{figure}

In this section, we evaluate the performance of \gls{curd} with varying forget set configurations: \emph{Forget-1\,\%}, \emph{Forget-5\,\%}, \emph{Forget-10\,\%}, and \emph{Forget-20\,\%}.
Performance metrics averaged across different data partitions, are detailed in Figure~\ref{fig:forgetset.volume.exp} for various unlearning epochs.

For both query and document removal tasks, increasing the ratio of the to-be-forgotten samples resulted in slower convergence for \gls{curd} on forget and substitute sets. The decline in $\mathbf{P}_\text{forget}$ and the improvement in $\mathbf{P}_\text{correct}$ both decelerate with larger forget sets, though they eventually stabilise near 0.00 and 1.00, respectively. 

Stability in performance on the retain set was observed across all to-be-forgotten sample ratios, with nearly identical retention curves approaching 1.00 as shown in Figure~\ref{fig:forgetset.volume.exp}.
The inference of \gls{curd} on unseen data (test dataset) displayed consistent initial performance across all sizes, with slight variations emerging in later epochs. Test performance was marginally better with smaller forget sets, indicating an inverse relationship between forget set size and test efficacy.

\subsection{Discussion}

\subsubsection{Key Findings}
The proposed method, \gls{curd}, outperforms several baseline unlearning methods in the corrective unranking task. While models like Amnesiac excel in forgetting but struggle to maintain performance on retain and test datasets, others like Retain and CF preserve performance on these sets but fail in effective forgetting. \Gls{curd} resolves this trade-off, delivering strong forgetting and correction performance while maintaining robust performance on the retain and test datasets.

%\axelin{The bold formatting of paragraph headings is inconsistent with the rest of the document}
\textbf{\gls{curd} is parameter-insensitive and adaptable to varying forget set sizes.} \Gls{curd} demonstrates parameter insensitivity, with its performance remaining stable despite variations in core parameters, such as the number of negative documents and quantile level. Moreover, \gls{curd} converges effectively across different forget set sizes.

\textbf{\gls{curd} enables controllable forgetting.}  As shown in Figure~\ref{fig:forgetset.volume.exp}, consistent with the concept of controllable forgetting proposed by \citet{hou2024neuralmachineunranking}, \gls{curd} maintains stable performance on retain and test datasets while achieving variable levels of forgetting performance.

\subsubsection{Implications}

\jerryRevise{\Gls{curd} is suited to retrieval systems that require fine-grained control over content removal and correction without sacrificing overall ranking quality. In academic search platforms, for instance, newly published, updated, or retracted documents must be handled efficiently. \gls{curd} enables local unranking adjustments in response to such changes without necessitating full model retraining, ensuring that system responsiveness and availability are maintained even under continuous updates. Moreover, CuRD's architecture supports the correction of erroneous or outdated associations in query-document relevance while preserving overall retrieval accuracy. }

\jerryRevise{From a compliance standpoint, \gls{curd}'s explicit modeling of forget and retain sets enhances transparency and auditability, which is valuable for systems that must comply with legal constraints such as the ``right to be forgotten''\citep{bourtoule2021machine}. However, deploying CuRD in practice also presents challenges, including the reliable identification of forget and correction instances.}

\subsubsection{Limitations}

\jerryRevise{\textbf{Synchronous forgetting and correcting.} One limitation of CuRD is that forgetting and correcting are performed synchronously. As shown in Figure~\ref{fig:forgetset.volume.exp}, improvements in correction accuracy are closely coupled with decreases in performance on the forget set, indicating that forgetting must proceed in tandem with correction. This tight coupling may hinder CuRD’s applicability in dynamic or real-time environments, where forgetting must occur immediately (e.g., following a user deletion request or regulatory takedown) but correction can be delayed or executed asynchronously. In such scenarios, the need for full retraining on both forget and substitute pairs may introduce latency or complicate deployment. Addressing this limitation could involve developing asynchronous or modular variants of CuRD that allow staged forgetting and later correction, or integrating continual learning mechanisms that enable incremental updates without synchronous processing.}

\jerryRevise{\textbf{Dependence on forget and retain sets.} CuRD relies on explicit access to both the forget set and the retain set to achieve effective correction and performance retention. This requirement limits its applicability in settings where historical training data is partially inaccessible or where data governance prohibits storing original samples. Compared to retraining-free approaches (e.g., SSD~\citep{Foster_Schoepf_Brintrup_2024}), this dependency may reduce operational flexibility. In particular, CuRD may be less suitable for edge devices or distributed systems with constrained memory and data access. A promising future direction is to develop synthetic-information-driven methods \citep{Zhang_Shen_Chen_Xu_2025}, enabling CuRD-style correction under stricter data availability constraints.}

\subsubsection{Future work}

\textbf{Designing corrective unranking methods for generative retrieval.} The current formalisation of corrective unranking is rooted in the learning-to-rank architecture, which may not fully align with the emerging \emph{generative retrieval} paradigms. Generative retrieval ranks documents directly in response to a query, leveraging the capabilities of generative language models~\citep{sun2024learning}. 
Future work could explore the development of corrective unranking methods specifically tailored for generative retrieval models.

\jerryRevise{\textbf{Integrating unlearning with continual learning for adaptive retrieval.} Future work could explore integrating machine unlearning with continual learning to enhance the adaptability of neural information retrieval systems. Continual learning allows neural \gls{ir} models to incrementally update with new data \cite{díazrodríguez2018dontforgetforgettingnew,HOU2025121368}, while machine unlearning focuses on removing specific information. 
This integration could enable \gls{ir} systems to retain new knowledge while selectively forgetting outdated or irrelevant data. This approach could be useful in dynamic environments where retrieval models must adapt to changing data distributions.}

\section{Conclusion} \label{sec:conclusion}

This work introduces the novel task of \emph{corrective unranking} within neural \gls{ir} systems, which not only involves forgetting undesirable data but also correcting retrieval outputs. We propose \gls{curd}, a new teacher--student framework designed to calibrate relevance scores for effective corrective unranking. Our experimental results show that \gls{curd} outperforms state-of-the-art baseline methods, achieving superior forgetting and correction performance while maintaining retention and generalisation abilities. Additionally, \gls{curd} is both parameter-insensitive and adaptable to varying sizes of the forget set, demonstrating robustness across different scenarios.

In summary, \gls{curd} enables neural IR systems to effectively forget user data while minimising the risks associated with data removal. \Gls{curd} has broad potential applications in various neural \gls{ir} contexts, including academic search platforms, digital archives, and news retrieval systems, where outdated information must be corrected while maintaining search quality. Additionally, \gls{curd} is valuable in e-commerce recommender systems by removing sensitive user information or erroneous product content without disrupting relevance.

\section*{Data availability}
 To reorganise the dataset and reproduce the experiments, please refer to the paper's GitHub repository located at
\href{https://github.com/JingruiHou/CorrectiveUnranking}{https://github.com/JingruiHou/CuRD}. 

\section*{Acknowledgments}
This work was supported by the China Postdoctoral Science Foundation under Grant No. 2025M773205, the Postdoctoral Fellowship Program of CPSF under Grant No. GZC20252319, and the National Natural Science Foundation of China under Grant No. 72374161. The authors would also like to thank the editors and anonymous reviewers for their valuable comments and suggestions, which greatly improved the quality of this manuscript.

\appendix

\section{Detailed corrective unranking result} \label{sec:detail.result}

This appendix provides a comprehensive analysis of the performance scores and unlearn time for seven baseline methods and our proposed method referred to as \gls{cocol}. We assess these eight corrective unranking methods using four neural ranking models on two datasets: MS MARCO and Trec CAR, with results detailed in Table~\ref{tab:unlearn_methods_marco} and Table~\ref{tab:unlearn_methods_trec} respectively.

In the evaluation of model performance on the retain set, a concern is that the performance score $\mathbf{P}_\text{retain}$ may not provide sufficient granularity to detect performance variations at the level of individual samples. Consider the following scenario: let $(0.5, 1, 0.5, 0.25)$ represent the reciprocal ranks for four queries $(\singleQuery_1, \singleQuery_2, \singleQuery_3, \singleQuery_4)$ in the retention set, processed by the trained model $\trainedModel$. After the application of corrective unranking, the unranking model $w$ yields reciprocal ranks of $(0.25, 0.5, 0.5, 1)$. Despite both models achieving identical performance scores, this metric fails to capture the nuanced shifts in individual query performance attributable to the unlearning process. Although no such shifts were observed in our experiments, it is necessary to define and report on such potential discrepancies for future investigations. To this end, we propose the following metric to quantify these shifts:
\begin{align}
    \mathbf{P}_{\Delta\text{retain}} & \coloneqq
\frac{1}{\lvert Q \rvert} \sum_{q \in Q} \frac{1}{\lvert D_q^+ \setminus D_q^f \rvert} \sum_{d \in D_q^+ \setminus D_q^f}  \Bigl( 
    \frac{1}{\rank_{w}(q, d; D_q^*)}
    \nonumber\\[-1ex]
    & \qquad\qquad\qquad\qquad - \frac{1}{\rank_{\trainedModel}(q, d; D_q)}
    \Bigr)^{\mathrlap{2}}.
\end{align}
Here, $D_q^*$ represents the set of to-be-ranked documents after correction, as defined in \eqref{eq:substitute_docs}.

\begin{table}[htbp]
    \caption{Performance metrics for various corrective unranking methods on MS MARCO. The symbol `$\uparrow$' indicates that higher scores are better, while `$\downarrow$' indicates that lower scores are better. }
    \centering
    \resizebox{\textwidth}{!}{
    \Large
    \begin{tabular}{llrrrrrrrr}
        \toprule
         Model & \makecell[l]{Unlearn \\ method} & {\makecell[r]{$\mathbf{P}_\text{forget}$ for \\ query \\removal ($\downarrow$)}} & {\makecell[r]{$\mathbf{P}_\text{forget}$ for \\ document \\ removal ($\downarrow$)}} & {\makecell[r]{$\mathbf{P}_\text{correct}$ for \\ query \\removal ($\uparrow$)}} & {\makecell[r]{$\mathbf{P}_\text{correct}$ for \\document\\ removal ($\uparrow$)}} & {$\mathbf{P}_\text{retain}$ ($\uparrow$)} & {$\mathbf{P}_{\Delta\text{retain}}$ ($\downarrow$)} & {$\mathbf{P}_\text{test}$ ($\uparrow$)} & \makecell[r]{Unlearn \\ time ($\downarrow$)} \\
        \midrule
        \multicolumn{1}{c}{\multirow{8}{*}{BERTcat}} & Retrain & 0.341 & 0.269 & 0.970 & 0.947 & 0.993 & 0.022 & 0.460 & 2.837 \\
 & CF & 0.512 & 0.448 & 0.955 & 0.944 & 1.000 & 0.018 & 0.479 & 14.298 \\
 & Amnesiac & 0.081 & 0.010 & 0.993 & 0.966 & 0.001 & 0.911 & 0.004 & 0.192 \\
 & NegGrad & 0.329 & 0.320 & 0.980 & 0.958 & 0.694 & 0.235 & 0.255 & 1.759 \\
 & BadT & 0.385 & 0.361 & 0.978 & 0.950 & 0.741 & 0.196 & 0.274 & 1.617 \\
 & SSD & 0.125 & 0.101 & 0.197 & 0.266 & 0.167 & 0.693 & 0.098 & 0.324 \\
 & CoCoL & 0.263 & 0.211 & 0.826 & 0.862 & 0.696 & 0.262 & 0.295 & 21.284 \\
 & Our & 0.022 & 0.020 & 0.957 & 0.932 & 0.987 & 0.021 & 0.436 & 1.524 \\ \midrule
\multirow{8}{*}{BERTdot} &  Retrain & 0.286 & 0.247 & 0.990 & 0.947 & 0.994 & 0.026 & 0.430 & 2.866 \\
 & CF & 0.439 & 0.374 & 0.960 & 0.934 & 0.998 & 0.041 & 0.536 & 14.342 \\
 & Amnesiac & 0.012 & 0.011 & 0.747 & 0.742 & 0.052 & 0.819 & 0.034 & 0.194 \\
 & NegGrad & 0.390 & 0.378 & 0.993 & 0.958 & 0.826 & 0.123 & 0.317 & 1.383 \\
 & BadT & 0.425 & 0.420 & 0.978 & 0.943 & 0.761 & 0.172 & 0.309 & 0.982 \\
 & SSD & 0.170 & 0.163 & 0.183 & 0.218 & 0.126 & 0.730 & 0.130 & 0.109 \\
 & CoCoL & 0.224 & 0.175 & 0.645 & 0.687 & 0.989 & 0.006 & 0.441 & 7.822 \\
 & Our & 0.020 & 0.017 & 0.987 & 0.969 & 0.989 & 0.018 & 0.396 & 0.547 \\ \midrule
\multirow{8}{*}{ColBERT} & Retrain & 0.475 & 0.403 & 0.899 & 0.887 & 0.961 & 0.037 & 0.554 & 2.887 \\
 & CF & 0.445 & 0.404 & 0.993 & 0.958 & 0.998 & 0.022 & 0.422 & 14.554 \\
 & Amnesiac & 0.030 & 0.010 & 0.960 & 0.943 & 0.040 & 0.792 & 0.033 & 0.195 \\
 & NegGrad & 0.246 & 0.232 & 0.978 & 0.949 & 0.498 & 0.354 & 0.315 & 2.190 \\
 & BadT & 0.303 & 0.281 & 0.978 & 0.949 & 0.524 & 0.334 & 0.330 & 1.211 \\
 & SSD & 0.134 & 0.113 & 0.285 & 0.330 & 0.286 & 0.544 & 0.233 & 0.333 \\
 & CoCoL & 0.271 & 0.220 & 0.591 & 0.615 & 0.948 & 0.025 & 0.625 & 8.923 \\
 & Our & 0.026 & 0.021 & 0.955 & 0.935 & 0.957 & 0.033 & 0.558 & 1.021 \\  \midrule
\multirow{8}{*}{PARADE}  & Retrain & 0.339 & 0.281 & 0.963 & 0.928 & 0.989 & 0.025 & 0.492 & 2.808 \\
 & CF & 0.482 & 0.432 & 0.970 & 0.942 & 0.999 & 0.024 & 0.520 & 14.080 \\
 & Amnesiac & 0.014 & 0.010 & 0.844 & 0.845 & 0.001 & 0.898 & 0.004 & 0.189 \\
 & NegGrad & 0.306 & 0.295 & 0.988 & 0.966 & 0.623 & 0.279 & 0.203 & 0.970 \\
 & BadT & 0.331 & 0.318 & 0.988 & 0.966 & 0.707 & 0.219 & 0.253 & 1.398 \\
 & SSD & 0.085 & 0.080 & 0.240 & 0.281 & 0.178 & 0.676 & 0.094 & 0.326 \\
 & CoCoL & 0.334 & 0.294 & 0.815 & 0.817 & 0.633 & 0.298 & 0.240 & 10.071 \\
 & Our & 0.015 & 0.012 & 0.964 & 0.937 & 0.974 & 0.031 & 0.467 & 2.192\\ \bottomrule
\end{tabular}%
}
\label{tab:unlearn_methods_marco}
\end{table}

\begin{table}[htbp]
    \caption{Performance metrics for various corrective unranking methods on Trec CAR. The symbol `$\uparrow$' indicates that higher scores are better, while `$\downarrow$' indicates that lower scores are better.}
    \centering
    \resizebox{\textwidth}{!}{
    \Large
    \begin{tabular}{llrrrrrrrr}
        \toprule
         Model & \makecell[l]{Unlearn \\ method} & {\makecell[r]{$\mathbf{P}_\text{forget}$ for \\ query \\removal ($\downarrow$)}} & {\makecell[r]{$\mathbf{P}_\text{forget}$ for \\ document \\ removal ($\downarrow$)}} & {\makecell[r]{$\mathbf{P}_\text{correct}$ for \\ query \\removal ($\uparrow$)}} & {\makecell[r]{$\mathbf{P}_\text{correct}$ for \\document\\ removal ($\uparrow$)}} & {$\mathbf{P}_\text{retain}$ ($\uparrow$)} & {$\mathbf{P}_{\Delta\text{retain}}$ ($\downarrow$)} & {$\mathbf{P}_\text{test}$ ($\uparrow$)} & \makecell[r]{Unlearn \\ time ($\downarrow$)} \\
        \midrule

\multirow{8}{*}{BERTcat} &  Retrain & 0.831 & 0.744 & 0.823 & 0.826 & 0.999 & 0.026 & 0.417 & 1.390 \\
 & CF & 0.975 & 0.875 & 0.755 & 0.755 & 1.000 & 0.012 & 0.505 & 6.070 \\
 & Amnesiac & 0.010 & 0.008 & 0.768 & 0.777 & 0.002 & 0.842 & 0.000 & 0.074 \\
 & NegGrad & 0.857 & 0.808 & 0.816 & 0.804 & 0.986 & 0.017 & 0.533 & 1.124 \\
 & BadT & 0.642 & 0.649 & 0.906 & 0.873 & 0.887 & 0.086 & 0.487 & 0.629 \\
 & SSD & 0.506 & 0.480 & 0.269 & 0.345 & 0.691 & 0.236 & 0.186 & 0.262 \\
 & CoCoL & 0.245 & 0.190 & 0.881 & 0.822 & 0.488 & 0.430 & 0.298 & 3.593 \\
 & Our & 0.076 & 0.028 & 0.873 & 0.863 & 0.987 & 0.033 & 0.456 & 0.632 \\\midrule
\multirow{8}{*}{BERTdot} &  Retrain & 0.864 & 0.779 & 0.814 & 0.795 & 0.996 & 0.033 & 0.408 & 1.339 \\
 & CF & 0.818 & 0.712 & 0.841 & 0.829 & 1.000 & 0.021 & 0.490 & 6.705 \\
 & Amnesiac & 0.021 & 0.018 & 0.251 & 0.326 & 0.095 & 0.722 & 0.057 & 0.071 \\
 & NegGrad & 0.843 & 0.785 & 0.824 & 0.813 & 0.993 & 0.019 & 0.449 & 1.082 \\
 & BadT & 0.782 & 0.729 & 0.842 & 0.833 & 0.918 & 0.056 & 0.489 & 0.608 \\
 & SSD & 0.217 & 0.129 & 0.222 & 0.283 & 0.219 & 0.604 & 0.031 & 0.249 \\
 & CoCoL & 0.365 & 0.331 & 0.289 & 0.346 & 0.997 & 0.007 & 0.300 & 27.335 \\
 & Our & 0.183 & 0.074 & 0.921 & 0.918 & 0.998 & 0.022 & 0.443 & 0.255 \\\midrule
\multirow{8}{*}{ColBERT} &Retrain & 0.905 & 0.807 & 0.769 & 0.756 & 0.991 & 0.034 & 0.487 & 1.357 \\
 & CF & 0.854 & 0.744 & 0.820 & 0.802 & 1.000 & 0.022 & 0.508 & 6.846 \\
 & Amnesiac & 0.015 & 0.012 & 0.287 & 0.350 & 0.106 & 0.713 & 0.040 & 0.072 \\
 & NegGrad & 0.782 & 0.755 & 0.830 & 0.814 & 0.863 & 0.099 & 0.492 & 1.095 \\
 & BadT & 0.842 & 0.785 & 0.820 & 0.815 & 0.998 & 0.012 & 0.472 & 0.616 \\
 & SSD & 0.285 & 0.248 & 0.206 & 0.281 & 0.248 & 0.584 & 0.120 & 0.251 \\
 & CoCoL & 0.409 & 0.297 & 0.307 & 0.373 & 0.998 & 0.004 & 0.493 & 17.100 \\
 & Our & 0.084 & 0.029 & 0.993 & 0.971 & 0.997 & 0.024 & 0.471 & 0.478  \\ \midrule
\multirow{8}{*}{PARADE} & Retrain & 0.792 & 0.703 & 0.829 & 0.843 & 0.998 & 0.027 & 0.489 & 1.288 \\
 & CF & 0.923 & 0.814 & 0.771 & 0.804 & 1.000 & 0.015 & 0.455 & 6.458 \\
 & Amnesiac & 0.011 & 0.008 & 0.990 & 0.947 & 0.039 & 0.776 & 0.010 & 0.073 \\
 & NegGrad & 0.720 & 0.724 & 0.838 & 0.834 & 0.922 & 0.070 & 0.455 & 1.118 \\
 & BadT & 0.628 & 0.606 & 0.887 & 0.881 & 0.869 & 0.104 & 0.442 & 0.647 \\
 & SSD & 0.079 & 0.050 & 0.235 & 0.339 & 0.086 & 0.733 & 0.000 & 0.145 \\
 & CoCoL & 0.541 & 0.384 & 0.812 & 0.816 & 0.879 & 0.107 & 0.413 & 28.887 \\
 & Our & 0.133 & 0.089 & 0.955 & 0.912 & 0.975 & 0.037 & 0.442 & 1.012 \\ \bottomrule
\end{tabular}%
}
    \label{tab:unlearn_methods_trec}
\end{table}

\jerryRevise{The comparative analysis of corrective unranking methods across the MS MARCO and TREC CAR datasets yields critical insights into the trade-offs among unlearning efficacy, model utility preservation, and computational efficiency.}

\jerryRevise{\textbf{Effectiveness of the proposed \gls{curd}.} Our proposed \gls{curd} method consistently demonstrates superior performance across multiple dimensions. In terms of forgetting efficacy, \gls{curd} achieves near-optimal $P_\text{forget}$ scores across nearly all model–dataset configurations, ranging from 0.015 to 0.183. These results indicate highly effective removal of target queries and documents. Importantly, \gls{curd} also delivers robust correction performance, with $P_\text{correct}$ values between 0.863 and 0.993,  outperforming baselines such as SSD and CoCoL. In retention performance, \gls{curd} shows minimal degradation, with $P_{\Delta\text{retain}}$ consistently falling between 0.018 and 0.037. This contrasts sharply with methods like Amnesiac and SSD, which severely damage retained model knowledge. Moreover, \gls{curd} exhibits competitive time efficiency, with unlearning times between 0.255 and 2.192, noticeably faster than methods like Retrain and CoCoL.}

\jerryRevise{\textbf{Limitations of baseline methods.} Among baseline approaches, Retrain remains a reliable yet computationally expensive method. While it maintains both correction and retention performance ($P_\text{correct}$ and $P_\text{retain} > 0.8$ on average), its forgetting ability is limited ($P_\text{forget} > 0.5$), and it is the slowest among all evaluated methods. Catastrophic Forgetting (CF) also fails to achieve effective forgetting ($P_\text{forget} > 0.6$  on average) despite prolonged training times. The Amnesiac approach, though aggressive in forgetting, severely compromises model utility, with $P_\text{retain}$ dropping below 0.1 and $P_\text{test}$ approaching zero—rendering the model practically unusable. SSD and CoCoL display severe trade-offs: SSD achieves modest forgetting but collapses in both correction and retention, with $P_\text{correct}$ falling below 0.35 on average; CoCoL, on the other hand, is both computationally intensive and unstable, exhibiting inconsistent correction performance with $P_\text{correct}$. NegGrad and BadT achieve moderate forgetting levels but incur considerable retain-set degradation, with $P_{\Delta\text{retain}}$ exceeding 0.1 in several settings.}

\jerryRevise{\textbf{Insights from the novel metric $P_{\Delta\text{retain}}$.} The newly proposed metric $P_{\Delta\text{retain}}$, which quantifies the degradation in retain-set performance. The proposed \gls{curd} achieves near-negligible degradation (0.018–0.037), suggesting that it effectively isolates the forget and correction signals through targeted parameter updates, without disrupting retained knowledge. In contrast, destructive baselines such as Amnesiac and SSD suffer catastrophic degradation, indicating that their unlearning procedures excessively perturb the model, sacrificing core functionality for aggressive forgetting. Moderate degradation is observed in NegGrad and BadT (0.151 and 0.144 on average). Although Retrain achieves excellent retention performance ($P_{\Delta\text{retain}}$ between 0.025 and 0.037), it does so at a substantial computational cost. Meanwhile, CF exhibits comparably low degradation (0.012–0.041), but fails to meet unlearning goals due to insufficient forgetting.}
%Bibliography
\bibliographystyle{unsrtnat}  
\bibliography{ref}

\end{document}